\documentclass[12pt]{article}

\usepackage{amsmath}
\usepackage{amssymb}
\usepackage{booktabs}  
\usepackage{color,soul}                     
\usepackage{graphicx}
\usepackage[round]{natbib}
\usepackage[margin={1in}]{geometry}
\renewcommand{\baselinestretch}{1.5}
\usepackage{algorithm}
\usepackage{algorithmic}
\usepackage{multirow}
\usepackage[table,xcdraw]{xcolor}
\usepackage{float}
\usepackage[section]{placeins}
\definecolor{brickred}{rgb}{0.6,0,0}
\RequirePackage
{hyperref} \hypersetup{
	colorlinks=true,       
	linkcolor=magenta,          
	citecolor=blue,        
	filecolor=magenta,      
	urlcolor=brickred           
}

\makeatletter

\def\spacingset#1{\renewcommand{\baselinestretch}%
	{#1}\small\normalsize} \spacingset{1}

\def\trans{^{\mathsf{T} }}
\newcommand{\rmd}{\mathrm{d}}

\newcommand{\bbE}{\mathbb{E}}

\newcommand{\bA}{{\mathbf{A}}}
\newcommand{\bB}{{\mathbf{B}}}

\newcommand{\bD}{{\mathbf{D}}}
\newcommand{\bF}{{\mathbf{F}}}
\newcommand{\bH}{{\mathbf{H}}}
\newcommand{\bI}{{\mathbf{I}}}

\newcommand{\bV}{{\mathbf{V}}}
\newcommand{\bP}{{\mathbf{P}}}
\newcommand{\bQ}{{\mathbf{Q}}}

\newcommand{\bS}{{\mathbf{S}}}
\newcommand{\bT}{{\mathbf{T}}}
\newcommand{\bZ}{{\mathbf{Z}}}
\newcommand{\bb}{{\mathbf{b}}}
\newcommand{\bc}{{\mathbf{c}}}

\newcommand{\bv}{\mathbf{v}}
\newcommand{\bz}{{\mathbf{z}}}

\newcommand{\balpha}{{\boldsymbol{\alpha}}}
\newcommand{\bbeta}{{\boldsymbol{\beta}}}
\newcommand{\btheta}{{\boldsymbol{\theta}}}

\newcommand{\bTheta}{{\boldsymbol{\Theta}}}
\newcommand{\bSigma}{{\boldsymbol{\Sigma}}}
\newcommand{\bGamma}{{\boldsymbol{\Gamma}}}

\newcommand{\blind}{1}

\begin{document}
\if1\blind
{
\title{Principal Component Analysis of Two-dimensional Functional Data with Serial Correlation}
\author{Shirun Shen\thanks{\baselineskip=10pt
		This article was majorly completed while Shirun Shen was a Ph.D. candidate in Department of Statistics, Texas A\&M University, College Station, TX 77843-3143.}~, Huiya Zhou, Kejun He\thanks{Corresponding author.}~, and Lan Zhou}  
\date{ }
\let\svthefootnote\thefootnote 
\let\thefootnote\relax\footnotetext{Shirun Shen (Email: \href{mailto:srshen@ruc.edu.cn}{srshen@ruc.edu.cn}) is Student, Huiya Zhou (Email: \href{mailto:freedom00y@ruc.edu.cn}{freedom00y@ruc.edu.cn}) is Student, Kejun He (Email: \href{mailto:kejunhe@ruc.edu.cn}{kejunhe@ruc.edu.cn}) is Assistant Professor, The Center for Applied Statistics, Institute of Statistics and Big Data, Renmin University of China, Beijing, 100872, China. 
Lan Zhou (Email:~\href{mailto:lzhou@stat.tamu.edu}{lzhou@stat.tamu.edu}) is Associate Professor, Department of Statistics, Texas A\&M University, College Station, TX 77843-3143. This research was supported by Public Computing Cloud, Renmin University of China.
}
\let\thefootnote\svthefootnote
\maketitle
} \fi 

\if0\blind
{
  \bigskip
  \bigskip
  \bigskip
  \title{Principal Component Analysis of Two-dimensional Functional Data with Serial Correlation}
  \author{}
  \date{}

  \maketitle
  \medskip
} \fi

\begin{abstract}

In this paper, we propose a novel model to analyze serially correlated two-dimensional functional data observed sparsely and irregularly on a domain which may not be a rectangle. Our approach employs a mixed effects model that specifies the principal component functions as bivariate splines on triangles and the principal component scores as random effects which follow an auto-regressive model. We apply the thin-plate penalty for regularizing the bivariate function estimation and develop an effective EM algorithm along with Kalman filter and smoother for calculating the penalized likelihood estimates of the parameters. Our approach was applied on simulated datasets and on Texas monthly average temperature data from January year 1915 to December year 2014.
\end{abstract}



\spacingset{1.46} 

\section{Introduction}\label{sec:intro}

Understanding the change of weather patterns over time and geological locations is important in studying the climate change. To investigate the temperature change in Texas, the United States, we worked on one data set from the U.S. Historical Climatology Network \citep[USHCN]{USHCN2009us}, collected by National Oceanic Atmospheric Administration (NOAA). This data set consists of monthly-average temperatures from year 1915 to year 2014, observed at sparsely located 49 weather stations in Texas.

\begin{figure}[h!]
	\centering
	\includegraphics[width=0.3\textwidth]{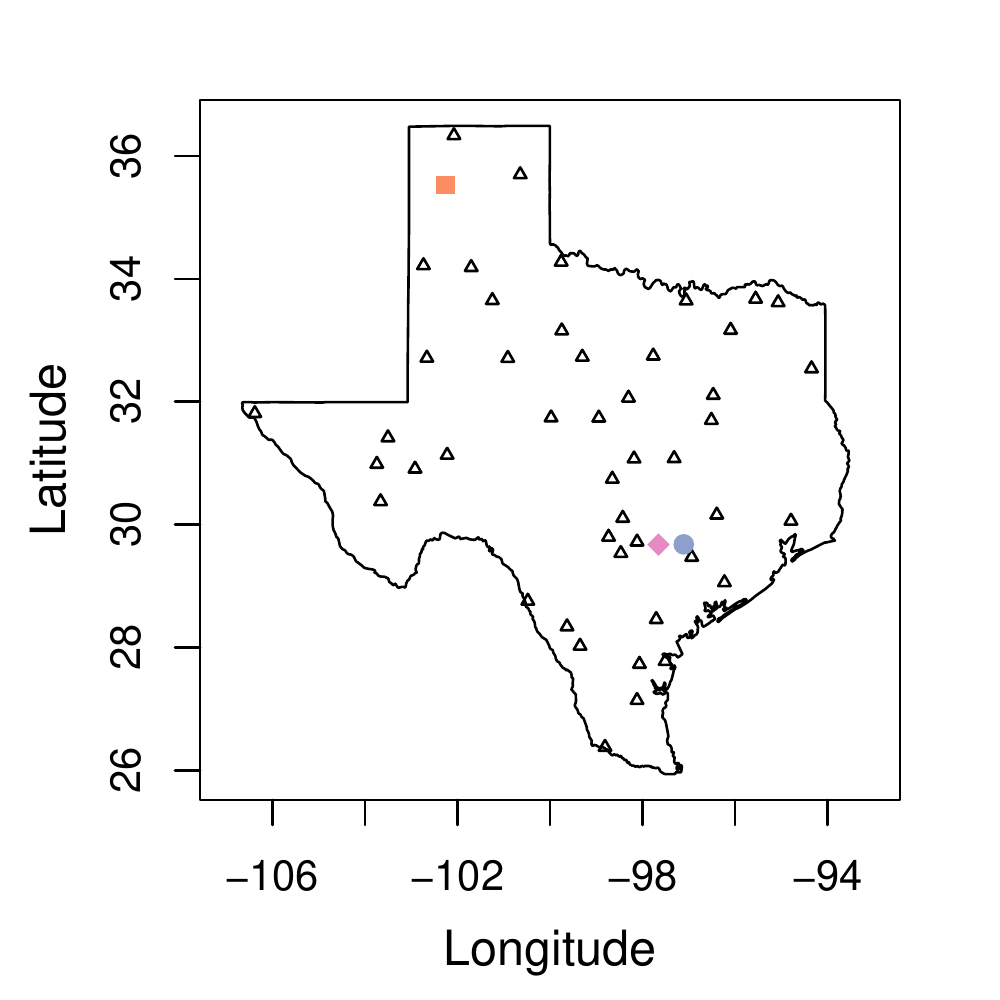}
    \includegraphics[width=0.3\textwidth]{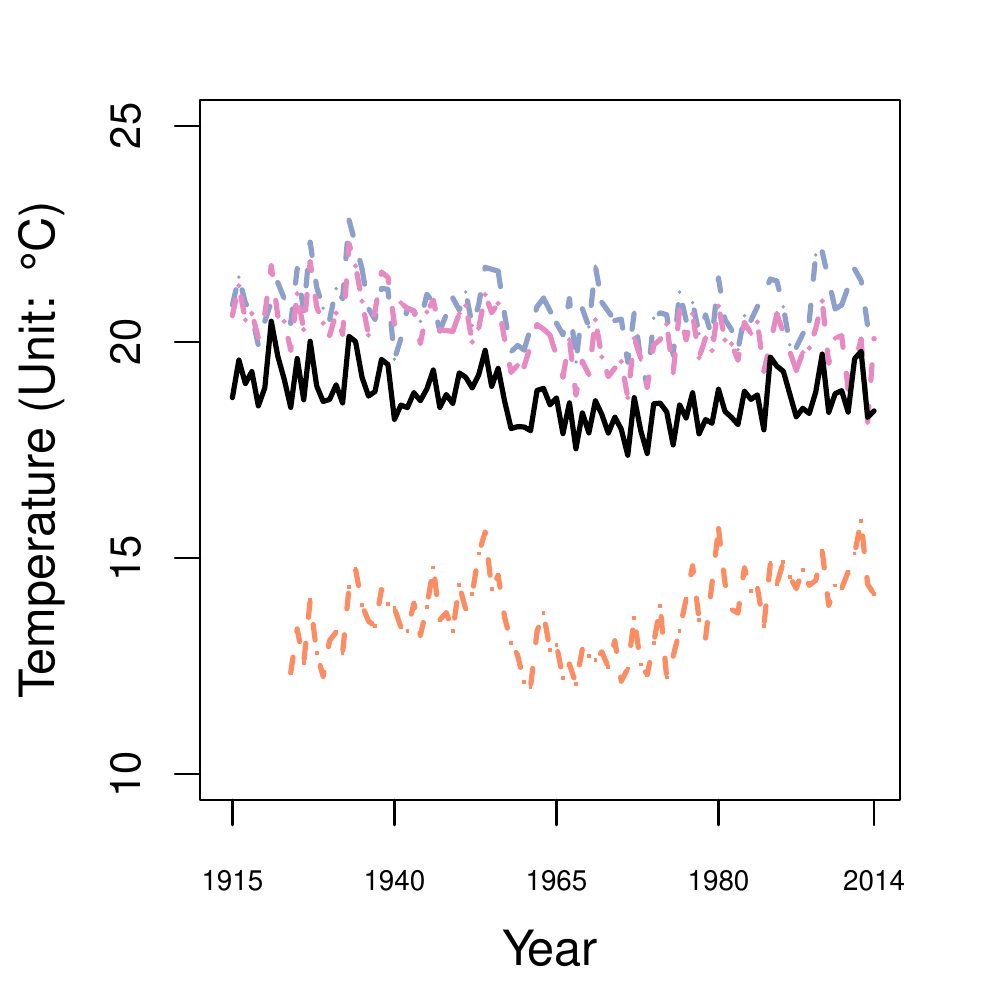}
    \includegraphics[width=0.3\textwidth]{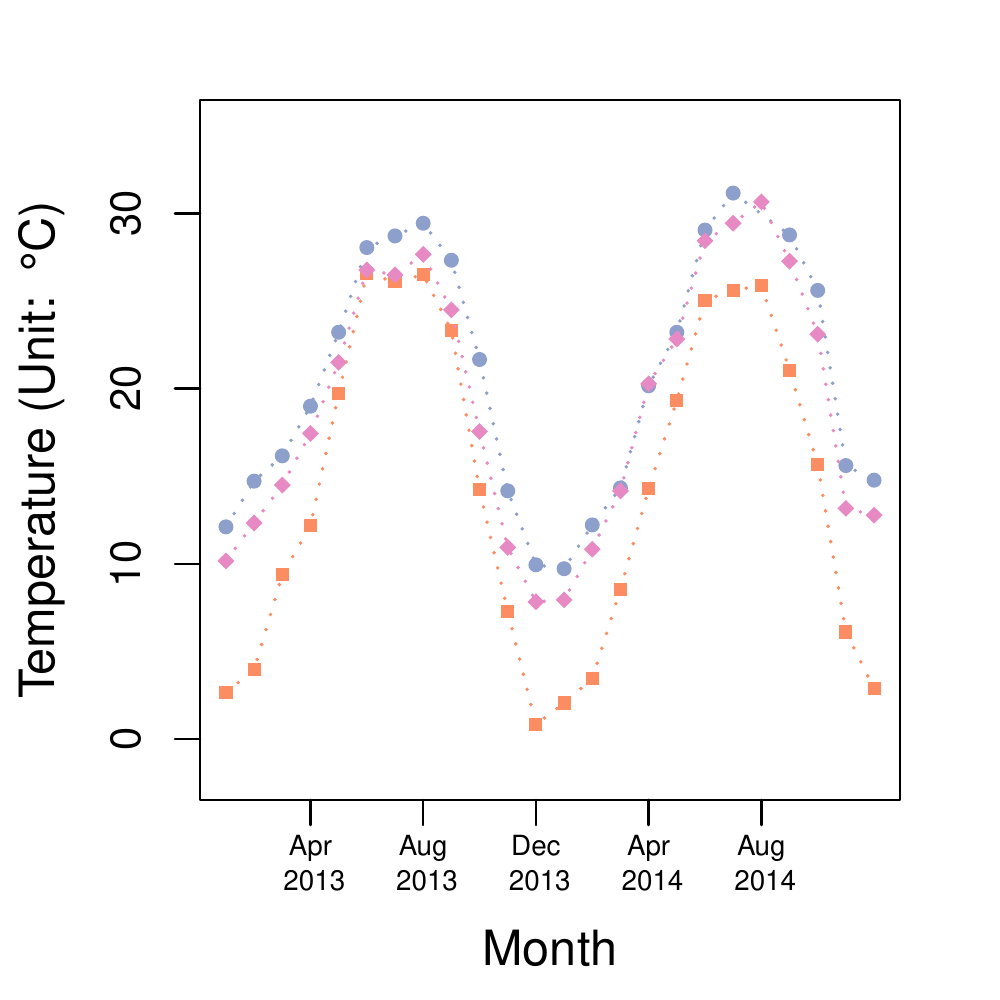}
	\caption{Left: Locations of $49$ weather stations in Texas. Middle: Yearly-average temperature (1915--2014) of $3$ weather stations (dashed lines) and the mean from all weather stations (solid black line). Right: Monthly average temperature of $3$ weather stations in year 2013 and 2014. In both middle and right panels, colors and symbols refer to the corresponding weather stations in the left panel. \label{fig:data_texas}}
\end{figure}

The left panel of Figure \ref{fig:data_texas} shows the locations of the $49$ weather stations in Texas. In particular, three weather stations are marked in different colors and symbols. 
The middle and right panels of Figure \ref{fig:data_texas} respectively give the yearly average temperatures of the three colored weather stations in the left panel from 1915 to 2014, and their monthly average temperatures in year 2013 and 2014. We can see that while the two neighboring southern weather stations follow similar trend over time, the northern weather station has lower temperatures consistently with different trend, especially before 1980. All three weather stations have almost identical seasons in year 2013 and 2014; compared with the two southern weather stations, the northern weather station's temperatures vary more between winters and summers. 
 
Studying the variation of weather patterns over the state of Texas across years using this data set provides the following challenges to data analysis. 
First, temperatures are affected by the locations of the measurement, as well as the trend of the climate change and the seasonality. 
Second, the weather stations are sparsely located in the irregular domain of Texas, and there are missing observations in various months from different weather stations. 
Third but not the least, there usually are serial correlations between temperatures observed over time; see for example, \cite{jones1986global} and \cite{hansen2006global}, especially when the temperatures are from the same weather station. 

To analyze temperature as a function of the location defined by the latitude and longitude, functional principal component analysis \citep[FPCA,][]{ramsay2006functional} provides an informative way of looking at the covariance structure of functional data that is otherwise not easily comprehensible.
The principal component functions represent the major mode of variation in functional data and characterize features of typical functions in a data set. 
When functional data are independent on a univariate domain, a rich literature exists in estimating a few leading functional principal component functions, including the local-polynomial methods \citep{staniswalis1998nonparametric,yao:muller:wang:05,yao2005functional,hall2006properties} and the spline methods \citep{rice2001nonparametric,james2000principal,huang2008functional,zhou2008joint,li2014functional,liu2017functional}. With 2-dimensional independent functional data, \cite{Zhou2014Principal} employed mixed effects models with bivariate splines on triangles to estimate the principal component functions. 
Other recent work on 2-dimensional functional data include
\cite{chen2017multi, wang2020simultaneous, ding2022functional,shi2022two}. None of these papers considered the serial correlations present in the data.


To analyze functional data with serial correlations, such as the Texas temperature data in our study, one approach is to model the functions directly by extending auto regressive models to functional data, see \cite{bosq2000linear} and \cite{kokoszka2013determining}. 
This approach, however, is not applicable when the functions are only sparsely observed as in the Texas temperature data.
Another approach with the idea of dimension reduction is a two-step procedure which first applies FPCA to get FPC scores, and then models the estimated FPC scores with time series models; see, for example, \cite{hyndman2007robust,shen2008interday,hyndman2009forecasting,shen2009modeling}, and references therein.
\cite{shang2017grouped} extended the work of \cite{hyndman2007robust} to deal with the data of multi-level structure.
\cite{cabrera2017forecasting} used penalized B-splines to construct the principal component functions for the generalized functional quantiles and, again, fit a time series model on the FPC scores. 
The aforementioned works, however, only considered one-dimensional functional time series, and cannot be simply extended to model 2-dimensional functional time series with missing data and observed on an irregular 2-dimensional domain, such as the Texas temperature data.

In this work, we propose a unified approach to model serially correlated 2-dimensional functional data and analyze the Texas temperature data with 
an FPCA model.
Specifically, to characterize the variation patterns of locations and overcome the challenges caused by sparse observations on the spatial domain as well as the irregular shape of Texas, following
\cite{Zhou2014Principal}, we
use bivariate splines on triangles to model the principal component functions. 
An auto-regressive (AR) model is employed on the FPC scores to capture the serial correlations over time. 
The trend of the climate change and the seasonality are respectively 
approximated by univariate splines and Fourier functions. 
All the above are integrated into a penalized complete data log likelihood function, and an EM algorithm is developed to estimate the unknown parameters. 
In the E-step of the EM algorithm, we further use Kalman filter and smoother to overcome the difficulty of calculating the expectations of  serially correlated latent variables.
Since an AR model is applied on the latent FPC scores,
our proposed serial functional principal component model (sFPC) does not require a minimum number of observations at each time point, and thus
can directly deal with missing observations as in the Texas temperature data.

The rest of this article is organized as follows. In Section~\ref{sec:model}, 
we develop a mixed effects model-based approach for principal component analysis of 2-dimensional functional data with serial correlation. We specify the principal component functions as bivariate splines on triangles and the principal component scores as random effects which follow an AR model. In Section~\ref{sec:modelestimation} we propose an EM algorithm for model fitting where Kalman filter and smoother procedures are applied in the E-step.
The empirical performance of the proposed method is illustrated via a simulation study and Texas temperature data analysis in Sections~\ref{sec:simu} and \ref{sec:realdata}, respectively. In Section~\ref{sec:summary}, we summarize the main contributions of this paper with some concluding remarks. 

\section{Mixed Effects Model for Serially Correlated 2-d Functional Data}\label{sec:model}
Let $\Omega$ be a compact subset of $\mathbb{R}^2$, and $(x,y)$ be the 2-dimensional index variable on $\Omega$. Suppose $Z(x,y)$ is a stochastic process on $\Omega$ with finite second moment, i.e., $\int_{\Omega} \mathbb{E}\{Z^2(x,y)\}\mathrm{d}x\mathrm{d}y < \infty$.
Denote the mean function of $Z(x,y)$ as $\mu(x,y) = \mathbb{E}\{Z(x,y)\}$ and the covariance function of $Z(x,y)$ as 
\[
	\mathcal{K}(x_1, y_1; x_2, y_2) = \mathbb{E}\big[ \{Z(x_1, y_1) - \mu(x_1,y_1)\} \{Z(x_2, y_2) - \mu(x_2,y_2)\} \big].
\]
Under mild conditions, Mercer's lemma \citep{mercer1909} shows that there exists an orthonormal sequence $\{\phi_j \}_{j}$ in $L_2(\Omega)$ as eigenfunctions, and a decreasing non-negative sequence $\{\zeta_j\}_j$ as eigenvalues, such that the covariance function can be expanded as
\[
	\mathcal{K}(x_1, y_1; x_2, y_2) = \sum_{j=1}^\infty \zeta_j \phi_j(x_1, y_1) \phi_j(x_2, y_2).
\]
The orthonormality of $\phi_j $'s means that $\int_{\Omega} \phi_j \phi_{j'}\,\rmd x \rmd y  = \delta_{jj'}$, where $\delta_{jj'}$ is the Kronecker delta.
Applying Karhunen-Lo{\`e}ve theorem \citep{karhunen1946spektraltheorie,loeve1946fonctions}, the random surface $Z(x,y)$ admits the following expansion
\begin{equation} \label{eqn:indepModel}
	Z(x,y) = \mu(x,y)+ \sum_{j=1}^{\infty}\alpha_j \phi_j(x,y),
\end{equation}
where $\alpha_j$'s are uncorrelated random variables with mean zero and variances $\{\zeta_j\}_j$.
Following \cite{ramsay2006functional}, the random variable $\alpha_j$ and the eigenfunction $\phi_j(x,y)$ are called the $j$-th FPC score and principal component function, respectively.

Assume that $Z(x,y)$ can be well approximated by its projection on the space spanned by the first $J$ eigenfunctions and treat the rest of terms as the noise, we arrive at the following model 
\begin{equation}\label{eqn:indepWorking}
	Z(x,y) = \mu(x,y)+ \sum_{j=1}^{J}\alpha_j \phi_j(x,y) + \epsilon(x,y),
\end{equation}
where $\epsilon(x,y)$ is a white noise with mean $0$ and variance $\sigma^2$.

When there are $n$ independent copies of $Z(x, y)$, denoted by $Z_1(x, y)$, $\dots$, $Z_n(x, y)$, \cite{Zhou2014Principal} proposed a mixed effects model-based approach and model the FPC scores as the random effects.
When the random surfaces $Z_t(x,y)$, $t= 1, \dots, n$, have time and location-dependent mean function $\mu_t(x,y)=\mathbb{E}\{Z_t(x,y)\}$ and $Z_t(x,y)$ are serially correlated, we need to consider both dependence between the locations and time points.

We first assume that location effect and the time effect are separable on the mean function such that $\mu_t(x,y) = \mu_1(x,y) \mu_2(t)$. Note that if one multiplies $\mu_1(x,y)$ by a non-zero constant $c$ and divides $\mu_2(t)$ by $c$, the value of $\mu_t(x,y)$ does not change. For identifiability purpose, we require that the $L_2$-norm of $\mu_1(x,y)$ 
be identity, i.e.,
\begin{equation}\label{eqn:identNorm}
\| \mu_1(x,y) \|_2 =1.
\end{equation}

Next borrowing the idea of FPCA as in \eqref{eqn:indepWorking}, 
we propose the model 
\begin{equation}\label{mixedeffects}
	Z_t(x,y) = \mu_1(x,y)\mu_2(t) + \sum_{j=1}^{J}\alpha_{j,t} \phi_j(x,y) + \epsilon_t(x,y), \quad t= 1, \dots, n,
\end{equation}
where $\alpha_{j,t}$ is the $j$-th FPC score at time $t$, $\phi_j(x,y)$ is the $j$-th principal component function which are orthonormal, 
and $\epsilon_t(x,y)$ is a white noise process with mean $0$ and variance $\sigma^2$. Furthermore, $\{\alpha_{j,t}\}_{t=1}^n$, $j=1, \dots, J$, are independent stationary time series. 
For each $j$, the time series, $\{\alpha_{j,t}\}_{t=1}^n$, follows the $p$-th order auto-regressive model (AR($p$)). To be specific, 
\begin{equation}\label{ARp}
    \alpha_{j,t} = \sum_{i=1}^p k_{i,j} \alpha_{j,t-i} + \eta_{j,t}, \quad \eta_{j,t} \overset{\rm ind}{\sim} \mathrm{N}(0, \sigma_{j}^2), \quad j=1,\dots, J \quad \text{and} \quad t= 1, \dots, n,
\end{equation}
where $k_{i,j}$'s and $\eta_{j,t}$'s are the auto-regressive coefficients and white noises of the AR($p$) models, respectively.
For the identifiability of the principal components, we assume that $\sigma_1^2>\dots>\sigma_J^2$.

Note that when $k_{i,j}=0$, $i=1, \dots, p$ and $j=1,\dots,J$, the FPC scores $\alpha_{j,t}$'s are mutually independent and normally distributed. Then the proposed model \eqref{mixedeffects} 
degenerates to the model in \cite{Zhou2014Principal}.


Assuming that the functions $\mu_1(x,y)$, $\mu_2(t)$ and $\phi_j(x,y)$ are smooth, we can approximate them by basis expansions. 
For the approximation of bivariate functions $\mu_1(x,y)$ and $\phi_j(x,y)$, we utilize the orthonormal bivariate Bernstein polynomials constructed on triangles \citep{lai2007spline} due to its advantage on irregular domains. 
The approximation properties of Bernstein polynomials were also theoretically studied in \cite{liu2016efficient}.
Please see Section \ref{sec:bivariate_triangulation} 
of Appendices for details of the bivariate basis on triangles.
As for the basis expansion of univariate function $\mu_2(t)$, we can choose from the commonly used regression spline \citep{de1978practical}, Bernstein spline \citep{lorentz2013bernstein}, and Fourier basis \citep{ramsay2006functional}, etc. 

Let $\mathbf{b}(x,y)$ denote an $n_b$-dimensional vector of orthonormal bivariate basis functions with
\begin{equation}\label{eqn:basisOrth}
	\int_{\Omega}\mathbf{b}(x,y)\mathbf{b}\trans(x,y) \rmd x \rmd y = \mathbf{I}_{n_b}, 
\end{equation}
and $\mathbf{c}(t)$ denote an $n_c$-dimensional vector of univariate basis functions. We write the basis expansions of the smooth functions as
\[
	\mu_1(x,y) = \mathbf{b}(x,y)\trans\boldsymbol{\theta}_b, ~~\mu_2 (t) = \mathbf{c}(t)\trans \boldsymbol{\theta}_c,
~~\mbox{and}~~ 
	\phi_j(x,y) = \mathbf{b}(x,y)\trans \boldsymbol{\theta}_j,\quad j=1,\dots, J,
\]
where the basis coefficients $\boldsymbol{\theta}_b \in \mathbb{R}^{n_b}$ with $\|\boldsymbol{\theta}_b\| = 1$, $\boldsymbol{\theta_c} \in \mathbb{R}^{n_c}$, and $\boldsymbol{\theta}_j \in \mathbb{R}^{n_b}$, $j=1,\dots, J$. In addition, $\btheta_j$, $j=1,\dots,J$ are orthonormal. Denote $\boldsymbol{\Theta}=(\boldsymbol{\theta}_1,\dots,\boldsymbol{\theta}_J) \in \mathbb{R}^{n_b \times J}$, $\balpha_t = (\alpha_{1,t},\dots,\alpha_{J,t})\trans$, and $\mathbf{K}_i = \mathrm{diag}(k_{i,1},\dots, k_{i,J})$, $i=1,\dots,J$, 
model \eqref{mixedeffects} can be rewritten as
\begin{equation}\label{eqn:themodel}
	Z_t(x,y) = \mathbf{b}(x,y)\trans\boldsymbol{\theta}_b\boldsymbol{\theta}_c\trans \mathbf{c}(t) + \mathbf{b}(x,y)\trans\boldsymbol{\Theta} \balpha_t + \epsilon_t(x,y), \quad t= 1, \dots, n,
\end{equation}
and the AR model \eqref{ARp} becomes
\[
	\balpha_t=\sum_{i=1}^p \mathbf{K}_i \balpha_{t-i} + \boldsymbol{\eta}_t, \quad t= 1, \dots, n,
\]
where $\boldsymbol{\eta}_t \sim \mathrm{N}(\mathbf{0},\mathbf{H}_J)$ with covariance matrix $\mathbf{H}_J = \mathrm{diag}(\sigma_1^2, \dots, \sigma_{J}^2)$. 

Suppose the sparsely sampled 2-dimensional surfaces are observed at time $t =1, \dots, n$. At time point $t$, there are $n_t$ randomly sampled points $(x_{t1},y_{t1}), \dots, (x_{tn_t},y_{tn_t})$ on the surface. Denote
$\mathbf{z}_t  = ( Z_t(x_{t1},y_{t1}),\dots,Z_t(x_{tn_t},y_{tn_t}) )\trans$, $\mathbf{B}_t = (\mathbf{b}(x_{t1},y_{t1}),\dots,\mathbf{b}(x_{tn_t},y_{tn_t}) )\trans$, $\boldsymbol{\epsilon}_t = (\epsilon_t(x_{t1},y_{t1}), \dots, \epsilon_t(x_{tn_t}, y_{tn_t}))\trans$, and $\mathbf{c}_t = \mathbf{c}(t)$ for notational simplicity. 
Model \eqref{eqn:themodel} for both observed data and latent variables can then be written as
\begin{equation}\label{models}
\begin{cases}
	\mathbf{z}_t = \mathbf{B}_t\boldsymbol{\theta}_b\boldsymbol{\theta}_c\trans\mathbf{c}_t+ \mathbf{B}_t \mathbf{\Theta}\balpha_t + \boldsymbol{\epsilon}_t, 
	\quad \boldsymbol{\epsilon}_t \sim \mathrm{N}(\mathbf{0},\sigma^2 \mathbf{I}_{n_t}),\\
	\balpha_t = \sum_{i=1}^p \mathbf{K}_i \balpha_{t-i} + \boldsymbol{\eta}_t,
	\quad \boldsymbol{\eta}_t \sim \mathrm{N}(\mathbf{0}, \mathbf{H}_J) ,
\end{cases}
\end{equation}
for $t= 1, \dots, n$, and the identifiability constraints are the same as mentioned above. Note that $n_t$ is the number of locations with observations at time $t$. It may vary with $t$, and is allowed to be 0. 

Denote $\mathbf{K} = (k_{i,j})_{i=1,j=1}^{p,J}\in \mathbb{R}^{p\times J}$. For model \eqref{models}, the unknown parameters to be estimated can be written collectively as $\Xi =  \{\boldsymbol{\theta}_b, \boldsymbol{\theta}_c, \mathbf{\Theta}, \mathbf{K}, \sigma^2, \{\sigma_{j}^2\}_{j=1}^J \}$.

\section{Model Fitting}\label{sec:modelestimation}
\subsection{Penalized Complete Data Log Likelihood}

Following model \eqref{models}, it is natural to estimate the unknown parameters $\Xi$ by maximizing the log likelihood function with some penalization to regularize the estimates of the smooth functions. However, since the latent FPC scores $\{\balpha_t\}_{t=1}^n$ follow an AR($p$) model, it is not feasible to integrate out $\{\balpha_t\}_{t=1}^n$ to get an analytical form of the log likelihood function of $\Xi$. 
By treating $\{\balpha_t\}_{t=1}^n$ as missing data, we can get an analytical form of the complete data log likelihood and then apply the EM algorithm \citep{dempster1977maximum} for parameter estimation. 

Let $\mathbf{k}_j$ be the $j$-th column of $\mathbf{K}$, i.e., $\mathbf{k}_j = (k_{1,j}, \dots, k_{p,j})\trans$, $j=1,\dots,J$. From model \eqref{models}, we derive the complete data log likelihood as
\allowdisplaybreaks
\begin{align}
	 -2 l_c(\Xi; \{\mathbf{z}_t\}_{t=1}^n, \{\balpha_t\}_{t=1}^n) 
	&= \frac{1}{\sigma^2}\sum_{t=1}^n (\mathbf{z}_t - \mathbf{B}_t \boldsymbol{\theta}_b \boldsymbol{\theta}_c\trans \mathbf{c}_t - \mathbf{B}_t \mathbf{\Theta}\balpha_t)\trans
	(\mathbf{z}_t - \mathbf{B}_t\boldsymbol{\theta}_b\boldsymbol{\theta}_c\trans \mathbf{c}_t  - \mathbf{B}_t \mathbf{\Theta}\balpha_t) \nonumber \\
	&\qquad + \sum_{t=1}^n n_t \log\sigma^2 + \sum_{j=1}^J\bigg\{ n\log \sigma_{j}^2 - \log \left|\mathbf{M}_{j}\right| + \frac{1}{\sigma_{j}^2} S_j(\mathbf{k}_j)\bigg\}, \label{completedatalik}
\end{align}
where $\mathbf{M}_{j}$ is the precision matrix of $(\alpha_{j,1}/\sigma_j,\dots, \alpha_{j,p}/\sigma_j)\trans$
and $S_j(\mathbf{k}_j)$ has the form of $(1,\mathbf{k}_j\trans) \mathbf{D}_j (1,\mathbf{k}_j\trans)\trans$ such that
\begin{equation}\label{eqn:sumSquareExpre}
	\mathbf{D}_j =
	\begin{pmatrix}
	D_{1,1,j} & -D_{1,2,j} & -D_{1,3,j} & \dots & -D_{1,p+1,j} \\
	-D_{1,2,j} & D_{2,2,j} & D_{2,3,j} & \dots & D_{2,p+1,j} \\
	\vdots & \vdots & \vdots & & \vdots \\
	-D_{p+1,1,j} & D_{p+1,2,j} & D_{p+1,3,j} & \dots & D_{p+1,p+1,j}
	\end{pmatrix} \in \mathbb{R}^{(p+1) \times (p+1)}
\end{equation}
with $D_{i,k,j} = D_{k,i,j} = \alpha_{j,i}\alpha_{j,k} + \alpha_{j,i+1}\alpha_{j,k+1} + \dots + \alpha_{j,n+1-k} \alpha_{j,n+1-i}$. 
The details of deriving \eqref{completedatalik} and \eqref{eqn:sumSquareExpre} are given in Section \ref{sec:completeLikeli} of Appendices.

We next introduce the roughness penalty for the regularization of the estimates of the smooth functions. For the univariate function $\mu_2(t)$, as in \cite{zhou2008joint}, we use the integrated squared second derivatives and the roughness penalty takes the form
\[
	\int_{T}\bigg\{\frac{\partial^2 \mu_2(t)}{\partial t^2}\bigg\}^2 \rmd t =  \btheta_c\trans \left[ \int_T \bigg\{ \frac{\partial^2 \bc(t)}{\partial t^2} \frac{\partial^2 \bc(t)\trans}{\partial t^2} \bigg\} \rmd t \right] \btheta_c := \btheta_c\trans \bP \btheta_c.
\]
For a generic bivariate function $f(x,y)$, we use the thin plate penalization \citep{ruppert2003semiparametric} which is defined as
\[
	\int_{\Omega} \Bigg[ \bigg\{ \frac{\partial^2 f(x,y)}{\partial x^2}\bigg\}^2 + 2 \bigg\{\frac{\partial^2 f(x,y)}{\partial x \partial y}\bigg\}^2 + \bigg\{ \frac{\partial^2 f(x,y)}{\partial y^2}\bigg\}^2 \Bigg] \rmd x \rmd y.
\]
Denote
\[
\bGamma = \int_{\Omega} \bigg\{ \frac{\partial^2 \bb(x,y)}{\partial x^2}\frac{\partial^2 \bb(x,y)\trans}{\partial x^2} + 2\frac{\partial^2 \bb(x,y)}{\partial x \partial y}\frac{\partial^2 \bb(x,y)\trans}{\partial x \partial y} + \frac{\partial^2 \bb(x,y)}{\partial y^2}\frac{\partial^2 \bb(x,y)\trans}{\partial y^2} \bigg\} \rmd x \rmd y.
\]
With basis expansions, the thin plate penalty for $\mu_1(x,y)$ and $\phi_j(x,y)$ can be written as, respectively, $\btheta_b \bGamma \btheta_b$ and $\btheta_j \bGamma \btheta_j$, $j =1 ,\dots ,J$.

Thus, the penalized complete data log likelihood has the expression
\begin{align}\label{penalizedlikelihood}
	& -2 l_c(\Xi; \{\mathbf{z}_t\}_{t=1}^n, \{\balpha_t\}_{t=1}^n) + \mathrm{Penalty}(\lambda ; \Xi) \nonumber \\
	&\qquad= \frac{1}{\sigma^2}\sum_{t=1}^n (\mathbf{z}_t - \mathbf{B}_t \boldsymbol{\theta}_b \boldsymbol{\theta}_c\trans \mathbf{c}_t - \mathbf{B}_t \mathbf{\Theta}\balpha_t)\trans
	(\mathbf{z}_t - \mathbf{B}_t\boldsymbol{\theta}_b\boldsymbol{\theta}_c\trans \mathbf{c}_t  - \mathbf{B}_t \mathbf{\Theta}\balpha_t)  + \sum_{t=1}^n n_t \log\sigma^2 \\
	&\qquad\qquad + \sum_{j=1}^J\bigg\{ n\log \sigma_{j}^2 - \log \left|\mathbf{M}_{j}\right| + \frac{1}{\sigma_{j}^2} S_j(\mathbf{k}_j)\bigg\} + \lambda_{\mu_s} \btheta_b\trans \bGamma \btheta_b + \lambda_{\mu_t} \btheta_c\trans \bP \btheta_c + \lambda_{pc}\sum_{j=1}^J \btheta_j\trans \bGamma \btheta_j,\nonumber
\end{align}
where $\lambda = (\lambda_{\mu_s}, \lambda_{\mu_t}, \lambda_{pc})$ are the regularization parameters. 

\subsection{EM Algorithm}\label{ssec:EM}

To estimate the parameters, instead of minimizing
\eqref{penalizedlikelihood}, the EM algorithm iteratively minimizes
\[
    Q(\Xi | \Xi^{(0)})= \mathbb{E}\big[ -2l_c(\Xi; \{\mathbf{z}_t\}_{t=1}^n, \{\balpha_t\}_{t=1}^n )  |\{\mathbf{z}_t\}_{t=1}^n, \Xi^{(0)}\big] + \mathrm{Penalty}(\lambda; \Xi),
\]
where $\Xi^{(0)}$ are the current guesses of the parameter values (i.e.values of the parameters from the previous iteration).

\paragraph{The E-step.} In the E-step, we calculate $Q(\Xi | \Xi^{(0)})$.  
Denote $\widehat{\balpha}_t = \mathbb{E}( \balpha_t| \{\mathbf{z}_t\}_{t=1}^n, \Xi^{(0)})$, $\widehat{\bSigma}_t  = \mathrm{Var}(\balpha_t|\{\mathbf{z}_t\}_{t=1}^n, \Xi^{(0)})$, and $\widehat{\mathbf{D}}_j = \mathbb{E}(\mathbf{D}_j|\{\bz_t\}_{t=1}^n, \Xi^{(0)})$, $j = 1, \ldots, J$. We obtain 
\begin{equation}\label{eqn:defSj}
	\widehat{S}_j(\mathbf{k}_j) :=\mathbb{E}[S_j(\mathbf{k}_j)| \{\mathbf{z}_t\}_{t=1}^n, \Xi^{(0)}] 
	= (1,\mathbf{k}_j)\trans \widehat{\mathbf{D}}_j (1,\mathbf{k}_j\trans)\trans.
\end{equation}
Using \eqref{completedatalik}--\eqref{eqn:defSj}, it shows that
\allowdisplaybreaks
\begin{align}\label{condlikelihoodwithpenalty}
	Q(\Xi | \Xi^{(0)}) &  = \frac{1}{\sigma^2}\sum_{t=1}^n\big\{(\mathbf{z}_t - \mathbf{B}_t\boldsymbol{\theta}_b\boldsymbol{\theta}_c\trans \mathbf{c}_t - \mathbf{B}_t \mathbf{\Theta}\widehat{\balpha}_t)\trans(\mathbf{z}_t - \mathbf{B}_t\boldsymbol{\theta}_b\boldsymbol{\theta}_c\trans \mathbf{c}_t - \mathbf{B}_t \mathbf{\Theta}\widehat{\balpha}_t) +  \mathrm{tr}(\mathbf{B}_t\mathbf{\Theta}\widehat{\bSigma}_t \mathbf{\Theta}\trans\mathbf{B}_t\trans)\big\} \nonumber \\
	& \qquad \qquad +  \sum_{t=1}^n n_t \log\sigma^2 +\sum_{j=1}^J\big\{ n \log \sigma_{j}^2 - \log \left|\mathbf{M}_{j}\right| + \frac{1}{\sigma_{j}^2} \widehat{S}_j(\mathbf{k}_j)\big\}   \\
	& \qquad \qquad 
	+ \lambda_{\mu_s} \btheta_{b}\trans \bGamma \btheta_b + \lambda_{\mu_t} \btheta_c \trans \bP \btheta_c + \lambda_{pc} \sum_{j=1}^J \btheta_j\trans \bGamma \btheta_j. \nonumber 
\end{align}
Hence, to calculate the value of \eqref{condlikelihoodwithpenalty}, we only need to calculate $\widehat{\balpha}_t$, $\widehat{\bSigma}_t$, and $\widehat{\mathbf{D}}_j$, $j =1, \ldots ,J$. 

Note that model \eqref{models} can be viewed as a state-space model \citep{durbin2012time}. 
To be specific, denote $\bbeta_t = (\balpha_{t+p}\trans, \dots, \balpha_t\trans)\trans$, $t = 1,\dots, n$, where $\balpha_{t} = 0$ when $t > n$. 
Denote $\mathbf{S} = \big(\mathbf{0},  \cdots  \mathbf{0} , \mathbf{I}  \big) \in \mathbb{R}^{J\times (p+1)J}$
and 
\[
	\mathbf{T}  =  \begin{pmatrix} \mathbf{K}_1 & \mathbf{K}_2 & \dots & \mathbf{K}_p & \mathbf{0} \\ \bI_J & \mathbf{0} & \dots & \mathbf{0} & \mathbf{0} \\ \mathbf{0} & \bI_J & \dots & \mathbf{0} & \mathbf{0} \\ \vdots & \vdots & & \vdots & \vdots \\ \mathbf{0} & \mathbf{0} & \dots & \bI_J & \mathbf{0}
	\end{pmatrix}
\in \mathbb{R}^{(p+1)J\times (p+1)J},
\]
model \eqref{models} can be rewritten as 
\begin{equation}\label{redef}
\begin{cases}
	\mathbf{z}_t = \mathbf{B}_t \btheta_b \btheta_c\trans \mathbf{c}_t + \mathbf{B}_t\mathbf{\Theta}\mathbf{S}\boldsymbol{\beta}_t + \boldsymbol{\epsilon}_t , \\
	\boldsymbol{\beta}_t = \mathbf{T}\boldsymbol{\beta}_{t-1} + 
	\boldsymbol{\xi}_t,
\end{cases}
\end{equation}
where $\boldsymbol{\xi}_t =
	(\boldsymbol{\eta}_{t+p}\trans, \mathbf{0}, \cdots , \mathbf{0}) \trans
	\sim \mathrm{N}(\mathbf{0}, \widetilde{\mathbf{H}})$ 
with $\widetilde{\mathbf{H}} =\mathrm{diag}(\mathbf{H}_J,\mathbf{0},\cdots, \mathbf{0}) \in \mathbb{R}^{(p+1)J \times (p+1)J}$.

Now with the state-space model \eqref{redef}, Kalman filter and smoother \citep{durbin2012time} can be used to obtain
$\widehat{\bb}_t = \bbE(\bbeta_t | \bz_1, \dots, \bz_n)$ and $\bV_t = \mathrm{Var}(\bbeta_t| \bz_1, \dots, \bz_n)$, and we can then get
\[
	\widehat{\balpha}_t = \mathbb{E}(\balpha_t | \bz_1,\dots,\bz_n, \Xi^{(0)} )= \mathbf{S}\widehat{\bb}_t 
	~~\mbox{and}~~
	\widehat{\bSigma}_t = \mathrm{Var}(\balpha_t|\bz_1,\dots,\bz_n, \Xi^{(0)} ) = \bS\bV_t \bS\trans.
\]
Following \eqref{eqn:sumSquareExpre}, $\widehat{\bD}_j$ can be obtained through computing $\widehat{D}_{i,k,j} := \mathbb{E}(D_{i,k,j}|\mathbf{Z},\Xi^{(0)})$ as
\begin{align*}
	\quad\widehat{D}_{i,k,j} & = \mathbb{E}(\alpha_{j,i}\alpha_{j,k} + \dots + \alpha_{j,n+1-k}\alpha_{j,n+1-i}|\mathbf{Z},\Xi^{(0)}) \\
	&= \mathbb{E}(\alpha_{j,i}|\bZ,\Xi^{(0)}) \mathbb{E}(\alpha_{j,k}|\bZ,\Xi^{(0)}) + \dots + \mathbb{E}(\alpha_{j,n+1-k}|\bZ,\Xi^{(0)}) \mathbb{E}(\alpha_{j,n+1-i}|\bZ,\Xi^{(0)}) \\
	& \qquad +\mathrm{Cov}(\alpha_{j,i},\alpha_{j,k}|\bZ,\Xi^{(0)}) + \dots + \mathrm{Cov}(\alpha_{j,n+1-k}, \alpha_{j,n+1-i}|\bZ,\Xi^{(0)}) \\
	&= \widehat{\alpha}_{j,i} \widehat{\alpha}_{j,k} + \dots \widehat{\alpha}_{j,n+1-k}\widehat{\alpha}_{j,n+1-i} +\sum_{t=1}^{n+1-i-k} \bV_{(1+p-i)J+j, (1+p-k)J+j,t}
\end{align*}
for $j =1, \ldots, J$, where $\widehat{\alpha}_{j,t}$ is the $j$-th element of $\widehat{\balpha}_t$ and $\bV_{(1+p-i)J+j, (1+p-k)J+j,t}$ is the $((1+p-i)J+j, (1+p-k)J+j)$-th element of $\bV_t$, $t=1, \dots,n$.

It remains to get $\widehat{\bb}_t$ and $\bV_t$ through Kalman filter and smoother \citep{durbin2012time}. The procedure of Kalman filter and smoother involves first applying Kalman filter algorithm then applying Kalman smoother in a reverse order. We give the details in the following paragraphs.
 
Let $\bb_{t|t-1}$ and $\bQ_{t|t-1}$ respectively be the one-step prediction of $\bbeta_{t}$ and its uncertainty, i.e., $\bb_{t|t-1} = \bbE\{\bbeta_t | \bz_{1:(t-1)},\Xi^{(0)}\}$ and $\bQ_{t|t-1} = \mathrm{Var}(\bbeta_t| \bz_{t-1},\dots, \bz_1,\Xi^{(0)})$, $t= 1,\dots,n$. Denote the estimation of the state at time $t$ and its uncertainty as $\bb_{t|t} = \bbE(\bbeta_t | \bz_t, \dots, \bz_1,\Xi^{(0)})$ and $\bQ_{t|t} = \mathrm{Var}(\bbeta_t| \bz_{t},\dots, \bz_1,\Xi^{(0)})$, respectively.
The Kalman filter operates in a prediction-correction loop. The prediction step updates
$$\begin{cases}
	\bb_{t|t-1}= \bT^{(0)}\bb_{t-1|t-1}\\ 
	\bQ_{t|t-1}= \bT^{(0)}\bQ_{t-1|t-1}(\bT^{(0)})\trans + \widetilde{\bH}^{(0)},
\end{cases} $$
where $\bT^{(0)}$ and $\widetilde{\bH}^{(0)}$ correspond to $\bT$ and $\widetilde{\bH}$ plugging in the current values of $\Xi^{(0)}$.
In the correction step, we denote $\bF_t = \bQ_{t|t-1}(\bB_t \bTheta^{(0)}\bS)\trans \{(\bB_t \bTheta^{(0)}\bS)\bQ_{t|t-1}(\bB_t\bTheta^{(0)}\bS)\trans + \sigma^{2(0)}\bI_{n_t} \}^{-1}$ as the Kalman gain matrix, and update
$$\begin{cases}
	\bb_{t|t} =\bb_{t|t-1} + \bF_t\{\bz_t - \bB_t\btheta_b^{(0)}(\btheta_c^{(0)})\trans \bc_t - 	\bB_t\bTheta^{(0)}\bS\bb_{t|t-1}\}  \\ 
	\bQ_{t|t} = \bQ_{t|t-1} - \bF_t(\bB_t \bTheta^{(0)}\bS) \bQ_{t|t-1}, 
\end{cases}$$
where $\bTheta^{(0)}$, $\sigma^{2(0)}$, $\btheta_b^{(0)}$, and $\btheta_c^{(0)}$ correspond to the current values of $\bTheta$, $\sigma^{2}$, $\btheta_b$, and $\btheta_c$, respectively. 
For the initialization, we adopt the commonly used non-informative values that $\bb_{0|0} = \mathbf{0}$ and $\bQ_{0|0} =\mathbf{0}$.

Next, we apply Kalman smoother recursion to obtain $\widehat{\bb}_t$ and $\bV_t$, $t=n-1, \dots, 1$, through the updating formula,
$$\begin{cases}
	\widehat{\bb}_t = \bb_{t|t} + \mathbf{L}_t(\widehat{\bb}_{t+1} - \bb_{t+1|t}) \\
	\bV_t = \bQ_{t|t} + \mathbf{L}_t (\bV_{t+1} - \bQ_{t+1|t})\mathbf{L}_t\trans,
\end{cases}$$
where $\mathbf{L}_t = \bQ_{t|t} (\bT^{(0)})\trans \bQ_{t+1|t}^{-1}$ with the initial values $\widehat{\bb}_n= \bb_{n|n}$ and $\bV_n = \bQ_{n|n}$ due to their definitions. 

We remark that the conditional expectations in the E-step can be calculated using the Kalman filter and smoother even when there are no observations at some time points; see, e.g.,  \cite{Cipra1997KalmanFW}.

\paragraph{The M-step.} In the M-step, we find the minimizer of $Q(\Xi | \Xi^{(0)})$ in \eqref{condlikelihoodwithpenalty}. The explicit form of minimizer is usually difficult to derive. 
Note that in \eqref{condlikelihoodwithpenalty}, the parameters, $\boldsymbol{\theta}_b, \boldsymbol{\theta}_c, \mathbf{\Theta}, \mathbf{K}, \sigma^2,$ and $\{\sigma_{j}^2\}_{j=1}^J$, are separated. 
Thus, alternatively, we use block-wise optimization and discuss the updating rules for each parameter when the others are fixed at the current values of $\Xi^{(0)}$.

The optimization problem with respect to $\boldsymbol{\theta}_b \in \mathbb{R}^{n_b}$ in \eqref{condlikelihoodwithpenalty}
is equivalent to minimizing  	
$(\boldsymbol{\theta}_b - \mathbf{m})\trans \mathbf{A}(\boldsymbol{\theta}_b - \mathbf{m})$, with the constraint that $\boldsymbol{\theta}_b\trans\boldsymbol{\theta}_b = 1$,
where
\begin{align*}
\begin{cases}
	\mathbf{m} = \big\{ \sum_{t=1}^n(\boldsymbol{\theta}_c^{(0)\mathsf{T}}\mathbf{c}_t)^2 \mathbf{B}_t\trans\mathbf{B}_t + \sigma^{2(0)} \lambda_{\mu_s}\bGamma \big\}^{-1} \sum_{t=1}^n(\boldsymbol{\theta}_c^{(0)\mathsf{T}}\mathbf{c}_t)\mathbf{B}_t\trans(\mathbf{z}_t - \mathbf{B}_t \mathbf{\Theta}^{(0)} \widehat{\balpha}_t) \\
	\mathbf{A} = \sum_{t=1}^n(\boldsymbol{\theta}_c^{(0)\mathsf{T}}\mathbf{c}_t)^2 \mathbf{B}_t\trans\mathbf{B}_t + \sigma^{2(0)} \lambda_{\mu_s}\bGamma.
\end{cases}
\end{align*}
By treating the sphere of $\boldsymbol{\theta}_b$ as an embedded sub-manifold of the $n_b$-dimensional Euclidean space, $\boldsymbol{\theta}_b$ can be updated using the gradient decent algorithm on a sub-manifold \citep{absil2009optimization}. 
Algorithm S.1 \ref{alg:armijo} in Section \ref{sec:detailGradSph} of Appendices specializes our implementation for using the gradient decent algorithm on the sphere sub-manifold.

We propose to update $\boldsymbol{\theta}_c$, $\sigma^2$, and $\sigma_{j}^2$, $j=1,\dots, J$,  by setting the corresponding block-wise derivatives to be zero. 
To be specific, by taking derivative of \eqref{condlikelihoodwithpenalty} with respect to $\boldsymbol{\theta}_c$ and setting it to zero, 
we update $\boldsymbol{\theta}_c$ by 
\[	
	\widehat{\boldsymbol{\theta}}_c = \bigg\{\sum_{t=1}^n(\mathbf{B}_t \widehat{\boldsymbol{\theta}}_b\mathbf{c}_t\trans)\trans(\mathbf{B}_t \widehat{\boldsymbol{\theta}}_b\mathbf{c}_t\trans) + \sigma^{2(0)} \lambda_{\mu_t} \bP\bigg\}^{-1}\sum_{t=1}^n(\mathbf{B}_t \widehat{\boldsymbol{\theta}}_b\mathbf{c}_t\trans)\trans(\mathbf{z}_t - \mathbf{B}_t \mathbf{\Theta}^{(0)}\widehat{\balpha}_t).
\]
Analogously, the updating formula for $\sigma^2$ is 
\[
\begin{aligned}
	\widehat{\sigma}^2 &= \frac{1}{\sum_{t=1}^n n_t} \sum_{t=1}^n \Big\{(\mathbf{z}_t - \mathbf{B}_t \widehat{\boldsymbol{\theta}}_b\widehat{\boldsymbol{\theta}}_c\trans \mathbf{c}_t - \mathbf{B}_t \mathbf{\Theta}^{(0)}\widehat{\balpha}_t)\trans(\mathbf{z}_t - \mathbf{B}_t\widehat{\boldsymbol{\theta}}_b\widehat{\boldsymbol{\theta}}_c\trans \mathbf{c}_t - \mathbf{B}_t \mathbf{\Theta}^{(0)}\widehat{\balpha}_t) \\
	& \qquad \qquad \qquad \qquad + \mathrm{tr}(\mathbf{B}_t\mathbf{\Theta}^{(0)}\widehat{\bSigma}_t \mathbf{\Theta}^{(0)\mathsf{T}}\mathbf{B}_t\trans)\Big\},
\end{aligned} 
\]
and the updating formula for $\sigma_{j}^2$ is $\widehat{\sigma}_{j}^2 = {\widehat{S}_j(\mathbf{k}^{(0)}_j)}/{n}$, $j =1,\dots, J$, where $\widehat{S}_j(\mathbf{k}^{(0)}_j)$ is as \eqref{eqn:defSj} with $\mathbf{k}^{(0)}_j$ (the $j$-th column of $\mathbf{K}^{(0)}$) plugged in.

For $\mathbf{\Theta}$, we first update the columns of $\mathbf{\Theta} = (\boldsymbol{\theta}_1,\dots,\boldsymbol{\theta}_J)$ sequentially. Minimizing \eqref{condlikelihoodwithpenalty} with respect to $\btheta_j$ is equivalent to minimizing
\[
	\sum_{t=1}^n \left\Vert \bz_t - \bB_t \widehat{\btheta}_b \widehat{\btheta}_c\trans \bc_t - \sum_{j' \neq j} \bB_t \btheta_{j'} \widehat{\alpha}_{j',t} - \bB_t \btheta_j \widehat{\alpha}_{j,t} \right\Vert^2 + \sum_{t=1}^n \mathrm{tr}(\bB_t \bTheta \widehat{\bSigma}_t\bTheta\trans \bB_t\trans) + \widehat{\sigma}^2 \lambda_{pc} \btheta_j\trans \bGamma \btheta_j,
\]
which has an analytical form
\[
	\widehat{\boldsymbol{\theta}}_j = \bigg\{\sum_{t=1}^n (\widehat{\alpha}_{j,t}^2 + \widehat{\bSigma}_{t,jj})\mathbf{B}_t\trans\mathbf{B}_t + \widehat{\sigma}^2\lambda_{pc}\bGamma\bigg\}^{-1} 
	\cdot \sum_{t=1}^n \mathbf{B}_t\trans \Big\{ (\mathbf{z}_t - \mathbf{B}_t\widehat{\boldsymbol{\theta}}_b\widehat{\boldsymbol{\theta}}_c\trans \mathbf{c}_t)\widehat{\alpha}_{j,t} - \sum_{j'\neq j}(\widehat{\alpha}_{j',t}\widehat{\alpha}_{j,t} + \widehat{\bSigma}_{t,j'j})\mathbf{B}_t\widehat{\boldsymbol{\theta}}_{j'} \Big\}. 
\]
To guarantee the orthonormality of $\widehat{\bTheta}$, we utilize the spectral decomposition of $\widehat{\bTheta}\widehat{\mathbf{H}}_J \widehat{\bTheta}\trans$, where $\widehat{\mathbf{H}}_J = \mathrm{diag}(\widehat{\sigma}_1^2, \dots, \widehat{\sigma}_J^2)$. 
In particular, let $\widehat{\bTheta}\widehat{\mathbf{H}}_J \widehat{\bTheta}\trans = \widetilde{\mathbf{Q}} \widetilde{\mathbf{D}} \widetilde{\mathbf{Q}}\trans$,
where $\widetilde{\mathbf{Q}}$ is orthonormal and $ \widetilde{\mathbf{D}}$ is a diagonal matrix with decreasing diagonal elements.
We then replace $\widehat{\bTheta}$ and $\widehat{\mathbf{H}}_J$ by $\widetilde{\mathbf{Q}}$ and $\widetilde{\mathbf{D}}$, respectively.
Furthermore, we replace $\widehat{\balpha}_t$ with $\widetilde{\mathbf{Q}}\trans \widehat{\mathbf{\Theta}} \widehat{\balpha}_t$, and such transformation preserves the variance of $\widehat{\mathbf{\Theta}} \widehat{\balpha}_t$.

Finally, we aim to minimize \eqref{condlikelihoodwithpenalty} with respect to $\mathbf{K}$, which is equivalent to minimizing
\begin{equation}\label{ols}
	\sum_{j=1}^J\bigg\{- \log \left|\mathbf{M}_{j}\right| + \frac{1}{\widehat{\sigma}_j^2} \widehat{S}_j(\mathbf{k}_j)\bigg\},
\end{equation}
where $\mathbf{M}_{j}$ is the precision matrix of $(\alpha_{j,1}/\sigma_j,\dots, \alpha_{j,p}/\sigma_j)\trans$ and $\widehat{S}_j(\mathbf{k}_j)$ is given in \eqref{eqn:defSj}.
Note that the value of $\log \left| \mathbf{M}_{j} \right|$ is invariant with the change of sample size $n$, while the second term in \eqref{ols} is $n$-dependent. To simplify the computation, we use the second term $\sum_{j=1}^J\big\{ ({1}/{\widehat{\sigma}_j^2}) \widehat{S}_j(\mathbf{k}_j)\big\}$
to approximate \eqref{ols}. 
Since $\widehat{S}_j(\mathbf{k}_j) = (1,\mathbf{k}_j\trans) \widehat{\mathbf{D}}_j (1,\mathbf{k}_j\trans)\trans$ has a quadratic form with respect to $\mathbf{k}_j$, using the weighted least squares, we can update $\mathbf{K}$ column-wise as 
$\widehat{\mathbf{k}}_j = \big( \widehat{\mathbf{D}}_{pj}\big)^{-1}\widehat{\mathbf{d}}_j$, $j=1,\dots, J$, where $\widehat{\mathbf{d}}_j = (\widehat{D}_{1,2,j}, \dots, \widehat{D}_{1,p+1,j})\trans$ and $\widehat{\mathbf{D}}_{pj}$ is the bottom right $p\times p$ major submatrix of $\widehat{\bD}_j$.

\subsection{Model Selection}\label{sec:modelselection}
The general guideline for constructing a triangulation is that we should avoid having triangles with a very small interior angle and that there is no triangle that contains no data point when the observed locations from all time points are pooled together.  
We refer to Chapter 4 of \cite{lai2007spline} for a detailed discussion of the triangulation. When the penalized spline method is used, the number of basis functions is not crucial in many applications as long as it is moderately large, since the roughness penalty helps regularize the estimation and prevent overfitting \citep{ruppert2003semiparametric}. Furthermore, the smoothness of the 2-dimensional basis functions $\bb(x,y)$ is determined by the order $d$ of polynomials and the order $r$ of the smoothness parameter on the connected edges of triangles. 
Practically, these two orders can also be given by the users based on the prior knowledge of the data. 
The number of PCs is determined by the empirical proportion of variances of temporal FPC scores. 
The order $p$ of auto-regressive model for the latent variables can be selected using a data-driven criteria like Akaike information criteria \citep[AIC,][]{akaike1974new} or Bayesian information criteria \citep[BIC,][]{schwarz1978estimating}, such that $p$ minimizes 
\[
    \sum_{j=1}^J\bigg\{ n \log \widehat\sigma_{j}^2  +     \frac{1}{\widehat\sigma_{j}^2} \widehat{S}_j(\widehat{\mathbf{k}}_j)\bigg\} + 2 p
~~\mbox{or}~~
    \sum_{j=1}^J\bigg\{ n \log \widehat\sigma_{j}^2  + \frac{1}{\widehat\sigma_{j}^2} \widehat{S}_j(\widehat{\mathbf{k}}_j)\bigg\} + \log(n) p. 
\] 
The regularization parameters $\lambda_{\mu_s}$, $\lambda_{\mu_t}$, and $\lambda_{pc}$ can be determined by minimizing the value of $K$-fold leave-location-out cross validation (CV), 
where the sample locations at each time point are randomly split into $K$ equal-sized subsets.
Nevertheless, since there are three regularization parameters, the classical full grid-search will be impractical due to high computational cost. Alternatively, we propose to use the simplex method \citep{Nelder1965simplex} to find the local optima. 
The overall selection procedure for the regularization parameters $(\lambda_{\mu_s},\lambda_{\mu_t}, \lambda_{pc})$ contains two steps. In the first step, we assign a few number of grid points sparse enough to cover a wide range of regularization parameters, and apply the $K$-fold CV to determine the best candidate according to the crossed predictive errors. Afterwards, we treat the selected point as the initial value and use the simplex method to obtain the final selected regularization parameters with local optimality.

\section{Simulation Study}\label{sec:simu}
In this section, we used a simulation study to compare the performance of the proposed serially correlated functional principal component (sFPC) model with that of the multivariate model (mFPC) in \cite{Zhou2014Principal} where the serial correlation is not considered. 

\begin{figure}[!h]
	\centering
	\includegraphics[width=0.4\textwidth]{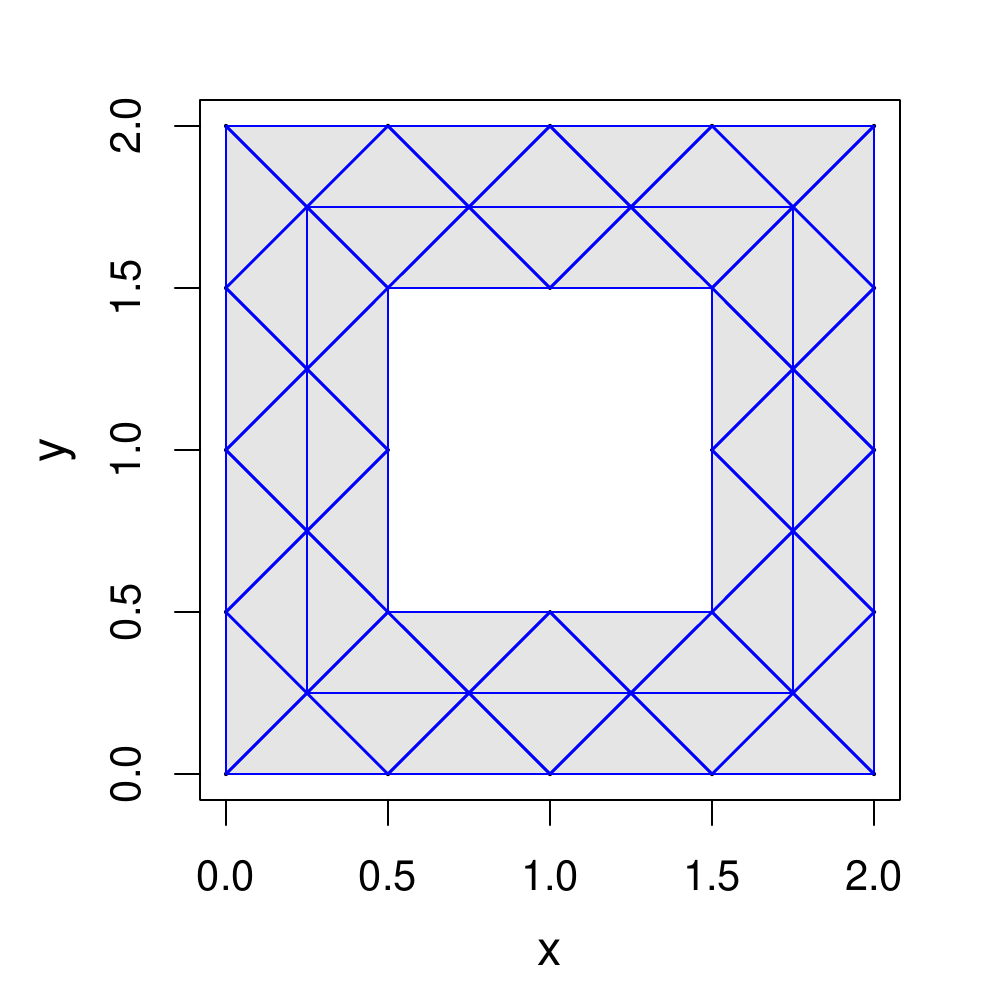}
	\vspace{-20pt}
	\caption{Domain $\Omega$ (with gray color) and its triangulation.
		\label{fig:tri_sim}}
\end{figure}

For the simulation experiment, the irregular domain $\Omega$ was set to be a $2 \times 2$ square with a hole of $1 \times 1$ square in the middle, as shown in Figure \ref{fig:tri_sim}. We generated data $Z_t(x,y)$ on $\Omega$ according to models \eqref{mixedeffects} and \eqref{ARp} with $J =2$ principal components and order of the AR model $p=2$. The mean function, $\mu_t(x,y)= \mu_1(x,y) \mu_2(t)$, and the principal component functions are given as follows, 
\begin{align*}
	\mu_1(x,y) &= 	5 \big\{\exp\big(\sqrt{0.1x^2 + 0.2y}\big) + \exp\big(-\sqrt{0.1x^2 + 0.2y}\big)\big\} , \\
	\mu_2(t) &= \cos(2\pi t/12) + t/n~~\text{or}~~ \mu_2(t) = 1, \\
	\phi_1(x,y) &= 0.8578\sin(x^2+0.5y^2), \\
	\phi_2(x,y) &= 0.8721\sin(0.3x^2+0.6y^2) - 0.2988 \sin(x^2+0.5y^2).
\end{align*}
Note that the principal component functions are orthonormal such that 
$\int_{\Omega} \phi_1^2(x,y) \rmd x \rmd y =1$, $\int_{\Omega} \phi_2^2(x,y) \rmd x \rmd y =1$, and $\int_{\Omega} \phi_1(x,y) \phi_2(x,y) \rmd x \rmd y =0$.

In the simulation study, we considered four setups: i) $\mu_2(t) = \cos(2\pi t/12) + t/n$ with AR(2) coefficients $k_{1,1}=k_{1,2}=0.8$ and $k_{2,1}=k_{2,2} = 0.1$; ii) $\mu_2(t) = \cos(2\pi t/12) + t/n$ with AR(2) coefficients $k_{1,1}=k_{1,2} = k_{2,1}=k_{2,2} = 0$; iii) $\mu_2(t) = 1$ with AR(2) coefficients $k_{1,1}=k_{1,2}=0.8$ and $k_{2,1}=k_{2,2} = 0.1$; iv) $\mu_2(t)=1$ with AR(2) coefficients $k_{1,1}=k_{1,2} = k_{2,1}=k_{2,2} = 0$. Note that in setups (ii) and (iv), the AR(2) model degenerates to a white noise model such that FPC scores are independent.
In each setup, we used two levels of variances: $\sigma^2=1, (\sigma_1^2, \sigma_2^2) = (1, 0.1)$; or $\sigma^2 = 0.1, (\sigma_1^2,\sigma_2^2) = (0.1, 0.01)$.
To simulate a data set, we set the number of time points $n=500$. 
At each time $t$, $t=1, \cdots, 500$, the number of observed locations was drawn from $\{50, 51,\dots,60\}$ uniformly and each location was randomly drawn from a uniform distribution on the irregular domain. 
We ran the simulation 100 times for each combination of the four setups and the two levels of variances. 
Both the proposed sFPC model 
and the mFPC model were applied on each simulated data.

\begin{table}[!h]
\centering
\begin{tabular}{cccccc}
\hline
    setup     & $\sigma^2$, $(\sigma_1^2,\sigma_2^2)$ & Model    & PA  & MIAE $\mu_t(x,y)$  & MIAE $Z_t(x,y)$   \\ \hline
    &  & sFPC & 4.6283 (0.0759) & 0.1001 (0.0052) & 0.1388 (0.0003) \\ 
    & \multirow{-2}{*}{1.0, (1.0, 0.1)}   & mFPC & 6.6644 (0.2103)  & 0.2223 (0.0068) & 0.1833 (0.0010) \\ \cline{2-6} 
    &  & sFPC & 4.6259 (0.0756) & 0.0324 (0.0017) & 0.0436 (0.0001) \\ 
    \multirow{-4}{*}{i)}   & \multirow{-2}{*}{0.1, (0.1, 0.01)}  & mFPC & 6.9333 (0.2605)  & 0.0771 (0.0027)  & 0.0585 (0.0003) \\ \hline
    &  & sFPC & 9.6592 (0.1411) & 0.0376 (0.0009) & 0.1384 (0.0003) \\ 
    & \multirow{-2}{*}{1.0, (1.0, 0.1)}   & mFPC                         & 9.7682 (0.1659)  & 0.0713 (0.0009) & 0.1498 (0.0003) \\ \cline{2-6}
    &  & sFPC & 9.7135 (0.1419) & 0.0126 (0.0003) & 0.0439 (0.0001) \\ 
    \multirow{-4}{*}{ii)}  & \multirow{-2}{*}{0.1, (0.1, 0.01)}  & mFPC & 11.908 (0.5291)  & 0.0264 (0.0003)  & 0.0487 (0.0002) \\ \hline
    &  &  sFPC & 6.9663 (0.0872) & 0.1052 (0.0045) & 0.1418 (0.0003) \\ 
    & \multirow{-2}{*}{1.0, (1.0, 0.1)}   & mFPC & 7.0707 (0.0884) & 0.0851 (0.0045)  & 0.1538 (0.0003) \\ \cline{2-6} 
    &  &  sFPC & 4.5682 (0.0751) & 0.0414 (0.0024) & 0.0430 (0.0001) \\ 
    \multirow{-4}{*}{iii)} & \multirow{-2}{*}{0.1, (0.1, 0.01)}  & mFPC  & 4.6599 (0.0771)  & 0.0502 (0.0030)  & 0.0473 (0.0001)  \\ \hline
    &  &  sFPC & 13.593 (0.1100) & 0.0316 (0.0012) & 0.1430 (0.0003) \\ 
    & \multirow{-2}{*}{1.0, (1.0, 0.1)}   & mFPC  & 13.582 (0.1092)  & 0.0302 (0.0010)  & 0.1425 (0.0003) \\ \cline{2-6} 
    &  &  sFPC & 8.9945 (0.1497) & 0.0110 (0.0004) & 0.0434 (0.0001) \\ 
    \multirow{-4}{*}{iv)}  & \multirow{-2}{*}{0.1, (0.1, 0.01)}  & mFPC  & 8.9911 (0.1494)  & 0.0110 (0.0004)  & 0.0433 (0.0001) \\ \hline
\end{tabular}
\caption{The means and standard errors of PAs and MIAEs for the mean function $\mu_t(x,y)$ and the stochastic surface $Z_t(x,y)$. The results are based on 100 simulation runs. 
	\label{tab:functions}
}
\end{table}

For both sFPC and mFPC models, on the spatial domain $\Omega$, we used the same triangulation shown in Figure \ref{fig:tri_sim}. The bivariate basis functions were constructed from Bernstein basis polynomials with $d = 3$ (cubic order splines) and $r = 1$ (continuous first derivative across the connected edges), same as that in \cite{zhou2014smoothing}. See Section \ref{sec:bivariate_triangulation} of Appendices for details on basis construction. 

On the time domain, to estimate $\mu_2(t)$ with sFPC model, we used 14 basis functions, including cubic polynomial to model the trend and Fourier basis functions $\sin({2k\pi t}/{12})$, $\cos({2k\pi t}/{12})$, $k=1, \cdots, 5$, to model the seasonality.

Since the mFPC model assumes a constant mean function over $t$, to apply the mFPC model in setups i) and ii), we took a two-step approach: 1) estimated the mean function $\mu_t(x,y)$ through basis expansion, and 2) fit the residuals with the mFPC model. 
To construct the basis functions in step 1), we first ignored the location effect and constructed an estimate of overall time effect, denoted as $\Tilde\nu(t)$, by regressing the data on the 14 time domain basis functions as that used in sFPC model. Next we generated the basis functions by multiplying $\Tilde\nu(t)$ with the bivariate basis functions on spatial domain $\Omega$ and for the estimate of $\mu_1(x,y) \mu_2(t)$ followed by regressing data on these basis functions. 
In setups iii) and iv) mFPC models were directly applied on the simulated data since the true mean function is constant over $t$, which meets the assumption of the mFPC model. 

For simplicity, to fit the models in our simulation study, the number of PCs and the order of AR were chosen to be the same as the true ones. 
We selected the three penalty parameters $(\lambda_{\mu_s}, \lambda_{\mu_t}, \lambda_{pc})$ by minimizing the value of 5-fold leave-location-out CV with the simplex method as described in Section~\ref{sec:modelselection}. 

To quantitatively measure the performance of the estimation of the mean function and the stochastic surfaces, we use the mean integrated absolute errors (MIAE) defined as
\[
	\frac 1n \sum_{t=1}^n\int_{\Omega} | \mu_t(x,y) - \widehat{\mu}_t(x,y)| \rmd x \rmd y \quad \text{and} \quad \frac 1n \sum_{t=1}^n\int_{\Omega} | Z_t(x,y) - \widehat{Z}_t(x,y)| \rmd x \rmd y,
\]
where each integration is evaluated as a scaled sum over $1976$ grid points distributed evenly on the spatial domain (the grid points are $51\times 51$ points evenly distributed on the rectangle with those in the hole been removed). 

Since the principal component functions are restrictive to be unit norm and can only be identified up to a sign change ($\pm 1$), we used the principal angle (PA) to evaluate the performance of the estimations of the principal component functions. The principal angle (PA) was calculated as follows:
We first evaluated the principal component functions on $1976$ grid points evenly distributed over the domain and obtained two matrices
$\widehat {\mathbf{V}}$ and $\mathbf{V}$, corresponding to the estimated and the true principal component functions. We then computed the principal angle as 
$\mathrm{angle} = \cos^{-1}(\rho) \times 180/ \pi$
where $\rho$ is the minimum singular value of the matrix $\bQ_{\widehat{\mathbf{V}}}\trans \bQ_{\mathbf{V}}$ with $\bQ_{\widehat{\mathbf{V}}}$ and $\bQ_{\mathbf{V}}$ being the orthonormal matrices of the QR decomposition of $\widehat{\mathbf{V}}$ and $\mathbf{V}$, respectively \citep{golub2013matrix}.

The averages and standard errors of MIAEs of the mean function $\mu_t(x,y)$ and stochastic surfaces $z_t(x,y)$, and PAs of the principal component functions from 100 simulation runs are summarized in Table \ref{tab:functions}. 
It is clear from this table that the sFPC method respectively reduces MIAE of the mean function and stochastic surfaces by $55\%$--$58\%$ and by $24\%$--$26\%$ compared with the mFPC method in setup i), 
while the improvement of the sFPC model on the estimation accuracy of principal component functions is around $30\%$--$33\%$ in terms of PA.
In setup ii), the improvement of sFPC can be observed on the estimation of mean function and principal component functions with $47\%$--$52\%$ MIAE reduction and $1\%$--$18\%$ PA reduction, respectively.
The outperformance of the sFPC model over the mFPC model in the estimation of stochastic surfaces is mild with $8\%$--$10\%$ less MIAE.  
This is not surprised since setup ii) assumes nonexistence of serial correlations between the FPC scores which is in favor of mFPC.
In setup iii), the sFPC method has respectively $17\%$ less and $24\%$ more MIAE of estimating the mean function in two levels of variances, because we set  $\mu_2(t)=1$ in this setup which meets the assumption of the mFPC model on time effect while the sFPC model estimates this constant $\mu_2(t)$ with a nonparametric method. 
However, compared with the mFPC method, the sFPC method still improves  $8\%$--$9\%$ in the estimation of stochastic surfaces and reduces about 
$2\%$ in PA.
In setup iv), where the true model is in favor of mFPC, the two models have similar results in the estimation of mean function, principal components, and stochastic surfaces.

\section{Texas Temperature Data Analysis}\label{sec:realdata}
\begin{figure}[!h]
	\centering
	\includegraphics[width=0.4\textwidth]{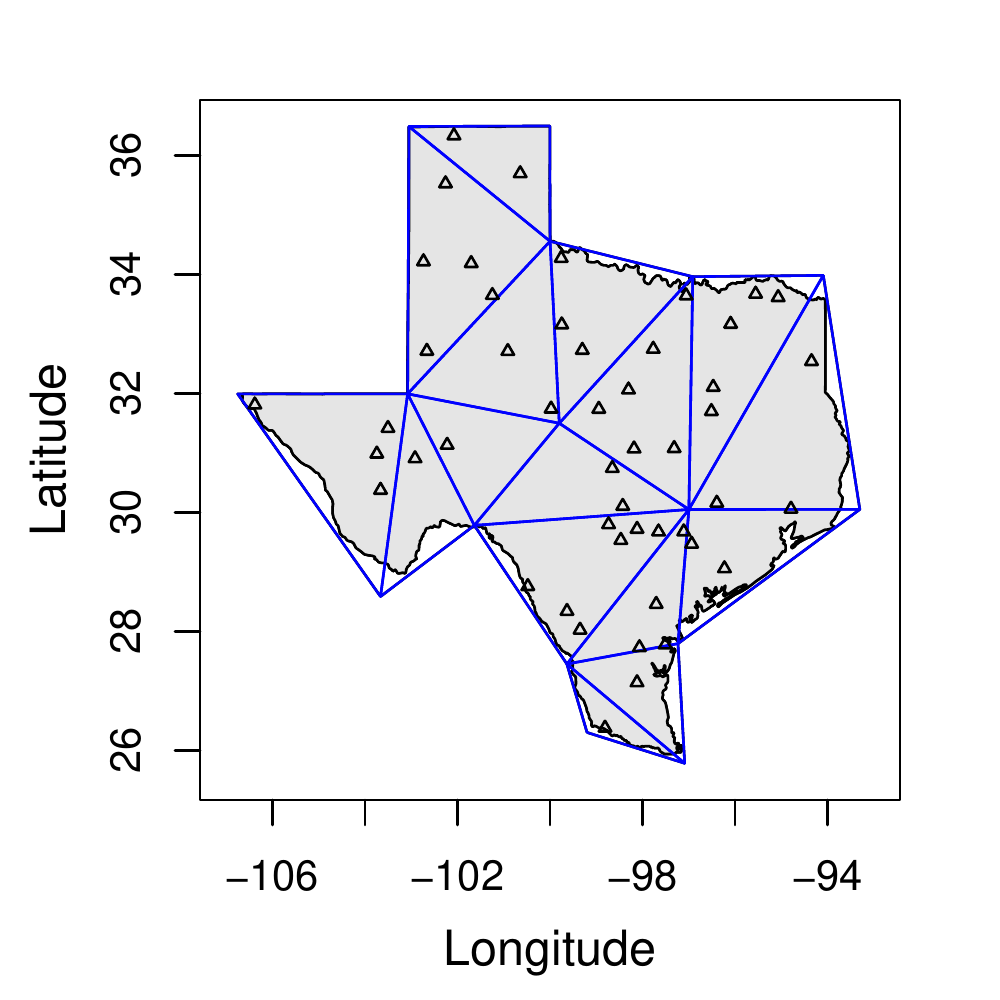}
	\vspace{-10pt}
	\caption{Triangulation used in the application to the Texas temperature data analysis.
		\label{fig:tri_real}}
\end{figure}

In this section, we apply the proposed model to study the climate change of Texas by analyzing Texas temperature data downloaded from United States Historical Climatology Network, Version 2.5 (USHCN v2.5, \url{https://cdiac.ess-dive.lbl.gov/epubs/ndp/ushcn/ushcn.html}). The data set consists of monthly average temperatures from January 1915 to December 2014 recorded by 49 weather stations located in Texas. The locations of these weather stations are shown in Figure \ref{fig:tri_real}.

With an area of 696,200 $\text{km}^2$, Texas has diverse physical geography and climate types. 
In general, in the eastern half of Texas, where lie the Gulf Coastal Plains and the North Central Plains, the climate is humid subtropical; in the western half, where lie the deserts and tall mountains, climate is semi-arid. 
Due to various reasons, $6.82\%$ of the data are missing. In particular, only $3$ stations have complete records while $13$ stations miss more than $120$ months of data.
There is no clear pattern in the missing of the data.

Let $W_t(x,y)$ denote the raw data at location $(x,y)$ and time $t$. Before applying the proposed sFPC model, we first estimated the location and time main effects by fitting a nonparametric regression for the raw data, i.e., $W_t(x,y) = \mu(x,y) + \nu(t) + \epsilon_t(x,y) = \bb(x,y)\trans \btheta_{\mu} + \bc(t)\trans \btheta_{\nu} + \epsilon_t(x,y)$, where $\bb(x,y)$ and $\bc(t)$ are basis functions on the location and time domains.
To be specific, the bivariate basis functions $\bb(x,y)$ were constructed from the Bernstein polynomials with $d=3$ and $r=1$ on the triangulation shown in Figure \ref{fig:tri_real}. This triangulation, also used in \cite{zhou2014smoothing}, covers the irregular domain of Texas with at least one weather station in each triangle. On the time domain, considering the seasonal effect and the climate change over time, we constructed the temporal basis functions $\bc(t) = \{\bc_1(t)\trans, \bc_2(t)\trans\}\trans$ where $\bc_1(t)$ are cubic splines with three interior knots at December of 1940, 1965, and 1990, to fit the trend, and $\bc_2(t)$ are Fourier functions $\sin( {2k\pi t}/{12})$, $\cos( {2k\pi t}/{12})$, $k=1, \cdots, 5$, to fit the seasonal effect. 
The penalized least squares approach was applied to estimate the parameters, and the penalty parameters were selected by 5-fold leave-location-out CV, where the sample locations at each time point are randomly split into $K$ equal-sized subsets.
Let $\widehat{\btheta}_{\mu}$ and $\widehat{\btheta}_{\nu}$ be the estimates of $\btheta_{\mu}$ and $\btheta_{\nu}$, respectively. 
The estimated location and time main effects are, respectively, $\widehat{\mu}(x,y) = \bb(x,y)\trans \widehat{\btheta}_{\mu}$ and $\widehat{\nu}(t) = \bc(t)\trans \widehat{\btheta}_{\nu}$.
Next we subtracted the estimated main effects from the raw data and obtained the demeaned data $Z_t(x,y) = 
W_t(x,y) - \widehat{\mu}(x,y) - \widehat{\nu}(t)$.

Following the sFPC method, $Z_t(x,y)$ were fitted with model \eqref{models} using the same basis functions $\bb(x,y)$ and $\bc(t)$ as described in the previous paragraph. As discussed in Section \ref{sec:modelselection}, we used $J = 3$ principal components after checking the empirical proportions of variances of temporal FPC scores and set the order of auto-regressive model to be $p=4$ which minimized the AIC. 
The three penalty parameters $\lambda_{\mu_s}, \lambda_{\mu_t}$, and $\lambda_{pc}$ were selected by 5-fold leave-location-out CV with simplex method as described in Section~\ref{sec:modelselection}. 
We also fitted $Z_t(x,y)$ with the mFPC method in \cite{Zhou2014Principal} for comparison. 
The mean absolute errors (MAEs) of the (in-sample) residuals for each month from both methods, presented in Table \ref{tab:fitting_MAE}, are similar such that the MAEs of the sFPC method are 94\% to 104\% of those of the mFPC method. 
However, the CV errors of sFPC and mFPC are respectively ${\rm CV}_{\rm sFPC} = 0.5679$ and ${\rm CV}_{\rm mFPC} = 1.999$. The smaller CV error of the proposed sFPC method demonstrates its better predictive performance on this real data.

\begin{table}[h!]
\centering
\resizebox{\textwidth}{!}{
\begin{tabular}{ccccccccccccc}
\hline
   & Jan  & Feb  & Mar  & Apr  & May  & Jun  & Jul  & Aug  & Sep  & Oct  & Nov  & Dec  \\ \hline
sFPC & 0.551 & 0.563 & 0.584 & 0.550 & 0.522 & 0.533 & 0.512 & 0.522 & 0.517 & 0.499 & 0.510 & 0.523 \\ 
mFPC & 0.571 & 0.568 & 0.560 & 0.570 & 0.536 & 0.548 & 0.520 & 0.525 & 0.517 & 0.509 & 0.526 & 0.556 \\ 
\hline
\end{tabular}
}
\caption{The MAEs of the residuals for each month from the sFPC and mFPC methods.}
\label{tab:fitting_MAE}
\end{table}

\begin{figure}[h!]
	\centering
    \includegraphics[width=0.9\textwidth]{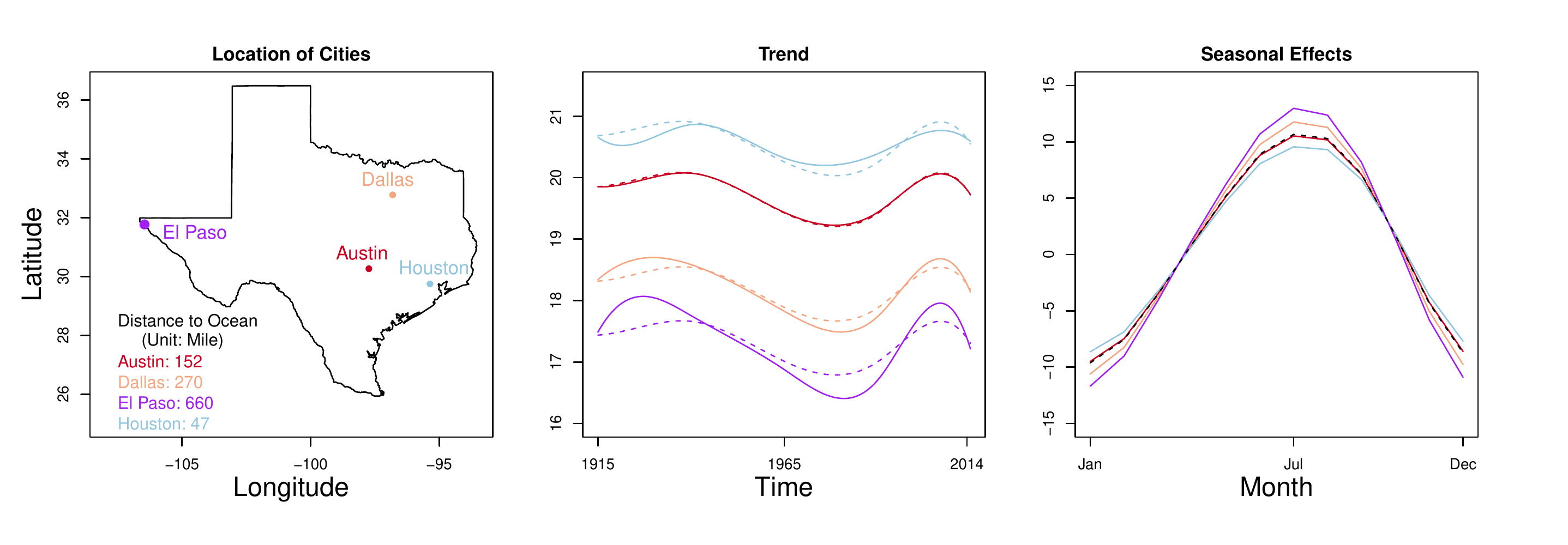}
	\caption{Left: The locations of  Austin, Dallas, El Paso, and Houston in Texas, and their distances to ocean.
	Middle: The estimated trends between January 1915 and December 2014 by two methods. Right: The estimated seasonal effects by two methods. 
	In the middle and right panels, the solid and dashed lines correspond to the estimated results using the sFPC and mFPC methods, respectively, and colors correspond to cities in the left panel.
	\label{fig:trendperiod}}
\end{figure}

Denote $\btheta_{\nu}=\{\btheta_{\nu 1}\trans, \btheta_{\nu 2}\trans\}\trans$ and $\btheta_{c}=\{\btheta_{c 1}\trans, \btheta_{c 2}\trans\}\trans$, where
$\btheta_{\nu 1}$ and $\btheta_{c 1}$ correspond to basis functions $\bc_1(t)$, and 
$\btheta_{\nu 2}$ and $\btheta_{c 2}$ correspond to basis functions $\bc_2(t)$. At Texas location $(x,y)$ and time $t$ between January 1915 December 2014, the estimated mean function
$\widehat{\mu}(x,y) + \widehat{\nu}(t) + \widehat{\mu}_1(x,y) \widehat{\mu}_2(t)$ by the sFPC method can be further decomposed as the estimated trend $\widehat{\mu}(x,y) +  \bc_1(t)\trans \widehat{\btheta}_{\nu 1} + \widehat{\mu}_1(x,y) \bc_1(t)\trans \widehat{\btheta}_{c 1}$ and 
the estimated seasonal effect 
$\bc_2(t)\trans \widehat{\btheta}_{\nu 2} + \widehat{\mu}_1(x,y) \bc_2(t)\trans \widehat{\btheta}_{c 2}$.
Figure \ref{fig:trendperiod} depicts the estimated trend and seasonal effects of Austin, Dallas, El Paso, and Houston between January 1915 and December 2014 as the solid lines for sFPC. It can be seen that all four cities have similar patterns in trend and seasonality. The temperatures are related to both latitude and altitude while the seasonal variation grows as the distance to the ocean increases. 
Similarly, we decomposed the estimated mean function by the mFPC method to the  trend and seasonal effects, and depicted in Figure \ref{fig:trendperiod} as the dashed lines.

\begin{figure}[h!]
	\centering
	\includegraphics[width=0.9\textwidth]{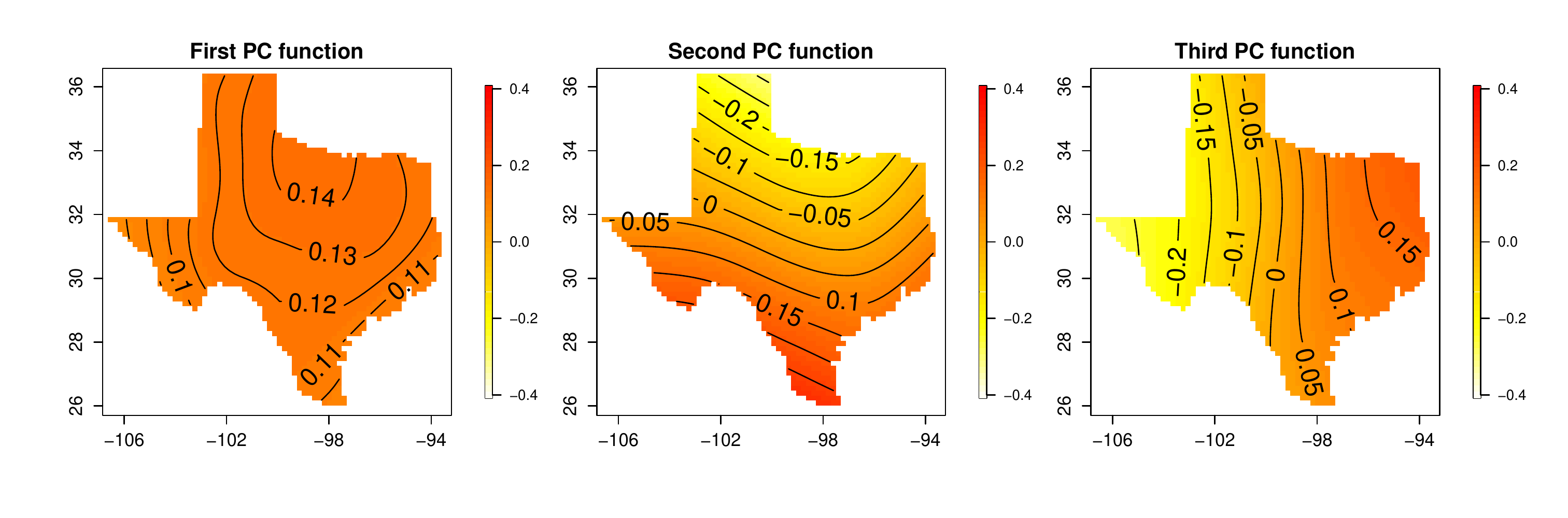}
	\includegraphics[width=0.9\textwidth]{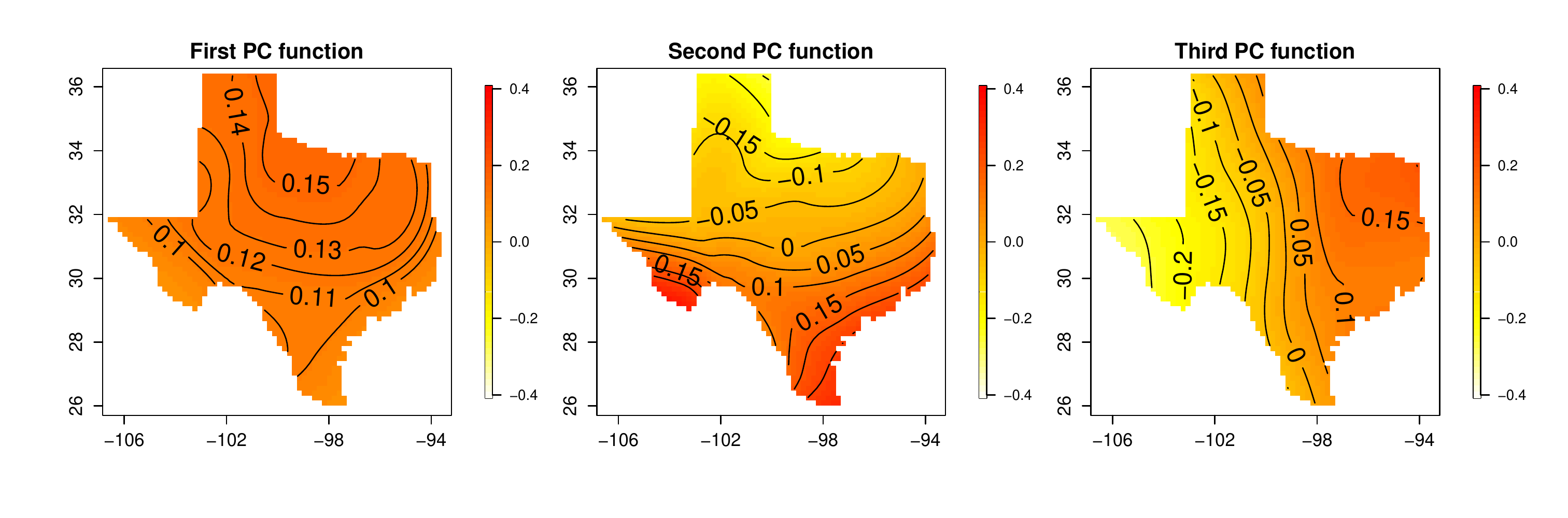}
	\caption{The first three estimated principal component functions using two methods for the real data analysis. The first and second rows respectively correspond to the sFPC and mFPC models. From left to right are the first, second, and third estimated principal component functions, respectively.
	\label{fig:pcfunc}}
\end{figure}

To show the main patterns of location variation estimated from the sFPC and mFPC method, we depict the heatmaps of the estimated 2-dimensional principal component functions (surfaces) with contours in Figure \ref{fig:pcfunc}. 
From both methods, the first estimated principal component surface varies has little variation over Texas, except in the area to the west of longitude $-104^\circ$, which may be an artifact due to the scarcity of the weather stations in that region. 
For the sFPC method, the contour lines of the second estimated principal component surface are roughly parallel to the latitude, and those of the third estimated principal component surface are roughly parallel to the longitude, which may be due to the changes in altitude.
For the mFPC method, the contour lines of the second estimated principal component surface may be affected by the Gulf of Mexico and North Central Plains, while the third estimated principal component surface is similar to that of sFPC.
We further used a semi-parametric bootstrap method to quantify the uncertainties of estimated PC functions. See Section \ref{sec:uncertain} of Appendices for the detailed procedure and the plots of bootstrap standard deviation surfaces.

We also compared the performance of short-term forecasting between sFPC and mFPC models by fitting models using data from 1915 to 2013 and predicting monthly average temperatures on year 2014. For both sFPC and mFPC, we fitted the data by following the same procedure as described in the previous paragraphs and used the same number of principal components ($J=3$) and the order of AR model ($p=4$). To predict the monthly temperature in 2014 with the sFPC model, we extrapolated the basis functions on the time domain for the mean functions and predicted the PC scores based on the fitted AR model. With the mFPC model, we used the estimated mean temperatures as the predicted values. Mean absolute prediction errors (MAPEs) of the 49 weather stations for the 12 months in 2014 are shown in Table \ref{tab:forecastRslt}.  Compared with the mFPC method, the sFPC method reduces the MAPEs of the prediction error by $2$\%--$39$\%, except in October where the sFPC method slightly increases the MAPE by $3$\%. 

\begin{table}[h!]
\centering
{\begin{tabular}{ccccccccccccc}
\hline
   & Jan  & Feb  & Mar  & Apr  & May  & Jun  & Jul  & Aug  & Sep  & Oct  & Nov  & Dec  \\ \hline
sFPC & 1.43 & 1.50 & 1.75 & 0.64 & 1.02 & 1.01 & 0.78 & 0.80 & 0.75 & 1.80 & 2.50 & 1.53 \\ 

mFPC & 1.74 & 1.80 & 1.88 & 0.74 & 1.25 & 1.37 & 1.27 & 1.11 & 0.82 & 1.74 & 2.55 & 1.60 \\ 
\hline
\end{tabular}
}
\caption{The mean absolute prediction errors of two methods for January--December, 2014.}
\label{tab:forecastRslt}
\end{table}

\section{Summary and Discussion}\label{sec:summary}
In this paper, we have proposed a novel approach, named sFPC, to analyze 2-dimensional functional data with serial correlations. In order to characterize the location and temporal effects of data, the bivariate splines on triangles and 
AR models on the latent FPC scores are jointly used.
An EM algorithm is developed with the Kalman filter and smoother to calculate the expected values in E-step.
With the latent AR models and EM algorithm, the proposed sFPC method can handle sparsity and irregularity on the surface domain, as well as missing observations on the time domain.
A simulation study and a real data analysis demonstrate that the sFPC method outperforms the mFPC method where the independence of the latent scores are assumed.

One limitation of the current model is that all noises are assumed to follow the Gaussian distribution. However, this assumption is often violated in many real problems. Thus one future research topic is to generalize our method by replacing the Gaussian distribution with other distributions such as the exponential family of distributions.
Moreover, it is restrictive to assume that the latent FPC scores of each principal component follow a stationary AR($p$) process. It is of interest to extend our method by considering a more general model where the latent FPC scores follow an ARIMA process, which can be non-stationary.

\section*{Ethics Declarations}
The authors have no conflicts of interest to declare.

\appendix 
\vspace{0.3in}
\noindent{\Large{\bf Appendices}}
\setcounter{section}{0}
\setcounter{equation}{0}
\setcounter{figure}{0}
\renewcommand{\thesection}{\Alph{section}}
\renewcommand{\theequation}{\Alph{section}.\arabic{equation}}
\renewcommand{\thefigure}{\Alph{section}.\arabic{figure}}
\renewcommand{\thealgorithm}{\Alph{section}.\arabic{algorithm}}

\section{Derivation of the Complete Data Log Likelihood}\label{sec:completeLikeli}
Treating the random effects $\balpha_i$'s as the latent variables, the negative twice log likelihood is
$$
	-2l_c(\Xi; \{\mathbf{z}_t\}_{t=1}^n, \{\balpha_t\}_{t=1}^n ) = -2\log p(\mathbf{z}_1,\dots,\mathbf{z}_n, \balpha_1,\dots,\balpha_n).
$$

The joint probability $p(\bz_1,\dots, \bz_n, \balpha_1,\dots, \balpha_n)$ can be decomposed into the multiple of the probability density $p(\balpha_1,\dots,\balpha_n)$ and the likelihood of observations given the latent variables, i.e., $p(\bz_1,\dots,\bz_n|\balpha_1,\dots,\balpha_n)$. Thus, the negative twice log likelihood can be written as
\begin{equation}\label{comlik}
\begin{aligned}
-2\log p (&\mathbf{z}_1,\dots,\mathbf{z}_n, \balpha_1,\dots,\balpha_n)  \\
&= -2\log p(\balpha_1,\dots,\balpha_n) -2 \log p(\mathbf{z}_1,\dots,\mathbf{z}_n | \balpha_1,\dots,\balpha_n),
\end{aligned}
\end{equation}
The AR($p$) time structure indicates that the first part in \eqref{comlik} can be decomposed as the joint density of initial states $\balpha_1, \dots, \balpha_p$ and the later states $\balpha_t | \balpha_{t-1}, \dots, \balpha_{t-p}$ for $t=p+1,\dots, n$
\begin{equation}\label{likdepart}
\begin{aligned}
	-2\log p(\balpha_1,\dots,\balpha_n) &= -2\log p(\balpha_1,\dots,\balpha_p) -2 \log p(\balpha_{p+1},\dots,\balpha_n |\balpha_1,\dots,\balpha_p) \\
	&=-2\log p(\balpha_1,\dots,\balpha_p) -2\sum_{t=p+1}^n \log p(\balpha_t | \balpha_{t-1},\dots,\balpha_{t-p}).
\end{aligned}
\end{equation}
The second term of \eqref{likdepart} has the explicit expression as
\[
\begin{aligned}
	&-2\sum_{t=p+1}^n \log p(\balpha_t | \balpha_{t-1},\dots,\balpha_{t-p})  \\
	& \qquad \qquad =\sum_{j=1}^J \sum_{t=p+1}^n \bigg\{ \log \sigma_{j}^2 + \frac{1}{\sigma_{j}^2}( \alpha_{j,t} - k_{1,j}\alpha_{j,t-1} - \dots - k_{p,j} \alpha_{j,t-p})^2 \bigg\},
\end{aligned}
\]
which comes from the conditional probability $\balpha_t | \balpha_{t-1},\dots, \balpha_{t-p} \sim \mathrm{N}(\sum_{i=1}^p \mathbf{K}_i \balpha_{t-i}, \mathbf{H}_J)$, where $\mathbf{K}_i = \mathrm{diag}(k_{i,1}, \dots, k_{i,J})$, $i=1,\dots,p$. 
According to \cite{box2015time}, the first term of \eqref{likdepart} can be derived as
\[
\begin{aligned}
	-2\log p(\balpha_1,\dots,\balpha_p) &= -2\sum_{j=1}^J\log p(\alpha_{j,1},\dots,\alpha_{j,p}) \\
 	&= \sum_{j=1}^J\bigg\{ p \log \sigma_j^2 - \log \left|\mathbf{M}_{j}\right| + \frac{1}{\sigma_j^2}S_{pj}(\mathbf{k}_j)\bigg\},
\end{aligned}
\]
where $\mathbf{M}_{j}$ is the precision matrix of $(\alpha_{j,1}/\sigma_j,\dots, \alpha_{j,p}/\sigma_j)\trans$ such that
\begin{equation}\label{eqn:precisionExpre}
	\mathbf{M}_{j} = 
	\begin{pmatrix}
	\gamma_{0,j} & \gamma_{1,j} & \dots & \gamma_{p-1,j} \\
	\gamma_{1,j} & \gamma_{0,j} & \dots & \gamma_{p-2,j} \\
	\vdots & \vdots & \ddots & \vdots \\
	\gamma_{p-1,j} & \gamma_{p-2,j} & \dots & \gamma_{0,j}
	\end{pmatrix}^{-1} \in \mathbb{R}^{p \times p}
\end{equation}
with $\gamma_{i,j} = \mathbb{E}(\alpha_{j,1+i}\alpha_{j,1})/ \sigma_j^2=\mathbb{E}(\alpha_{j,1}\alpha_{j,1+i})/\sigma_j^2$, $i=0, \dots, p-1$ and $j=1,\dots,J$ (depending on $\mathbf{k}_j = (k_{1,j},\dots, k_{p,j})\trans$); 
and $S_{pj}(\mathbf{k}_j) = \boldsymbol{\widetilde{\alpha}}_j\trans \mathbf{M}_{j}  \boldsymbol{\widetilde{\alpha}}_j$ 
is the residual sum of squares with $\boldsymbol{\widetilde{\alpha}}_j=(\alpha_{j,1},\dots, \alpha_{j,p})\trans$. 
Therefore, the first part of \eqref{comlik} is
\[
	-2\log p(\balpha_1,\dots,\balpha_n) = \sum_{j=1}^J\bigg\{ n \log \sigma_j^2 - \log \left|\mathbf{M}_{j}\right| + \frac{1}{\sigma_j^2} S_j(\mathbf{k}_j)\bigg\}, 
\]
where the total residual sum of squares is
\[
\begin{aligned}
	S_j(\mathbf{k}_j) &= S_{pj}(\mathbf{k}_j) + \sum_{t=p+1}^n ( \alpha_{j,t} - k_{1,j}\alpha_{j,t-1} - \dots - k_{p,j} \alpha_{j,t-p})^2 
	= (1,\mathbf{k}_j\trans)\trans \mathbf{D}_j (1,\mathbf{k}_j\trans),
\end{aligned}
\]
with 
\[
	\mathbf{D}_j =
	\begin{pmatrix}
	D_{1,1,j} & -D_{1,2,j} & -D_{1,3,j} & \dots & -D_{1,p+1,j} \\
	-D_{1,2,j} & D_{2,2,j} & D_{2,3,j} & \dots & D_{2,p+1,j} \\
	\vdots & \vdots & \vdots & & \vdots \\
	-D_{1,p+1,j} & D_{2,p+1,j} & D_{3,p+1,j} & \dots & D_{p+1,p+1,j}
	\end{pmatrix}, 
\]
and the $(i,k)$-th element $D_{i,k,j} = D_{k,i,j} = \alpha_{j,i}\alpha_{j,k} + \alpha_{j,i+1}\alpha_{j,k+1} + \dots + \alpha_{j,n+1-k} \alpha_{j,n+1-i}$. Further details can be referred to Appendix A7.4 of \cite{box2015time}. Following the relationship between observed variables $\mathbf{z}_t$ and latent variables $\balpha_t$, the second part of the complete data log likelihood \eqref{comlik}, i.e., the likelihood of the observed data given latent variables can be written as
\[
\begin{aligned}
	& -2\log \{p(\mathbf{z}_1,\dots, \mathbf{z}_n | \balpha_1, \dots, \balpha_n)\}\\ 
	& \qquad =  -2 \sum_{t=1}^n \log p(\mathbf{z}_t|\balpha_t)  \\
	& \qquad = \sum_{t=1}^n n_t \log\sigma^2 + \frac{1}{\sigma^2}\sum_{t=1}^n (\mathbf{z}_t - \mathbf{B}_t \boldsymbol{\theta}_b \boldsymbol{\theta}_c\trans \mathbf{c}_t - \mathbf{B}_t \mathbf{\Theta}\balpha_t)\trans
	(\mathbf{z}_t - \mathbf{B}_t\boldsymbol{\theta}_b\boldsymbol{\theta}_c\trans \mathbf{c}_t  - \mathbf{B}_t \mathbf{\Theta}\balpha_t). 
\end{aligned}
\]
Hence the complete data log likelihood can be written as in \eqref{completedatalik} of the main paper.

\section{Bivariate Spline Basis Functions on a Triangulation~\label{sec:bivariate_triangulation}}
In this section, we discuss the construction of the 2-dimensional orthonormal basis function $\mathbf{b}(x,y)$. One trivial choice is the tensor-product B spline basis functions, i.e., $\mathbf{b}(x,y) = \mathbf{b}_1(x) \otimes \mathbf{b}_2(y)$. However, the tensor-product B spline basis functions will cause two problems: (1) the computational cost is usually expensive due to a large number of tensor-product basis functions; and (2) this basis can only used in regular regions like a rectangle. To overcome these challenges, we alternatively introduce the bivariate Bernstein polynomials on triangulations. \cite{lai2007spline} presented the mathematical properties of the bivariate spline. \cite{Zhou2014Principal} applied such bivariate splines into their mFPC model. Due to the great approximation properties of Bernstein polynomials \citep{liu2016efficient}, there emerged many applications in spatial statistical models \citep{Yu2019Triangulation, Wang2019Triangulation}.
Figure \ref{fig:tri_sim} and Figure \ref{fig:tri_real} of the main paper depict the triangulation examples that will be used in our simulation study and Texas temperature data analysis. As we can see, unlike the commonly used tensor product of univariate basis functions, this bivariate basis can easily handle the irregular shapes on $\mathbb{R}^2$.

Denote $\delta$ as a triangle, which has the counter-wise vertices $(\bv_1, \bv_2, \bv_3)$. Then for any point $\bv \in \mathbb{R}^2$, there is a unique representation in the form $\bv = b_1 \bv_1 + b_2\bv_2 + b_3 \bv_3$. The three coefficients $(b_1, b_2, b_3)$ are called the barycentric coordinates of $\bv$ with respect to the triangle $\delta$. Given a non-negative integer $d$ and for any $i,j,k$ such that $i+j+k=d$, the Bernstein polynomials of degree $d$ relative to triangle $\delta$ are defined as
\[B_{ijk}^{d}(\bv) = \frac{d!}{i!j!k!}b_1^i b_2^j b_3^k.\]

Let $P_d(\delta)$ be the space of polynomials defined on the triangle $\delta$ with degree $d$. Then the Bernstein polynomials $B_{ijk}^d, i+j+k=d$, form a basis for $P_d(\delta)$. That is, for any function $s \in P_d(\delta)$, we have
\[s(\bv) = \sum_{i+j+k=d} \gamma_{ijk} B_{ijk}^d(\bv).\]

For an irregular domain, we can construct a triangulation $\Delta = \{\delta_1, \dots, \delta_M\}$ whose union covers the irregular region $\Omega$ \citep[see, for example,][]{lai2007spline}. 
We construct the Bernstein polynomial basis functions with respect to each $\delta_i$, and the collection of all such polynomials form a basis for $P_d(\Delta)$, the space of continuous piecewise polynomials of degree $d$ on $\Delta$. With additional smoothness conditions that the derivatives up to $r$ degrees are continuous, the bivariate basis functions can be constructed accordingly. The details of imposing such smoothness condition and applying the Gram-Schmidt procedure to ensure the orthonormal condition of the basis functions $\bb(x,y)$ are referred to \cite{zhou2014smoothing}.

\setcounter{equation}{0}
\setcounter{figure}{0}
\section{Details of the M-step}\label{sec:MStepDetail}
In the M-step, we find the minimizer of $Q(\Xi | \Xi^{(0)})$ in \eqref{condlikelihoodwithpenalty} of the main paper. 
Note that in \eqref{condlikelihoodwithpenalty}, the parameters, $\boldsymbol{\theta}_b, \boldsymbol{\theta}_c, \mathbf{\Theta}, \mathbf{K}, \sigma^2,$ and $\{\sigma_{j}^2\}_{j=1}^J$, are separated. 
Thus, we propose to use block-wise optimization, i.e., in each iteration, we update each parameter with the others fixed at the current values.

\subsection{Updating $\boldsymbol{\theta}_b$}\label{sec:detailGradSph}
Recall that updating $\boldsymbol{\theta}_b$ is equivalent to finding the minimizer of the following objective function:
\begin{equation}\label{sphereopti}
    f(\boldsymbol{\theta}_b) = (\boldsymbol{\theta}_b - \mathbf{m})\trans \mathbf{A}(\boldsymbol{\theta}_b - \mathbf{m}),
\end{equation}
with the constraint that $\boldsymbol{\theta}_b\trans\boldsymbol{\theta}_b = 1$,
where
\[
	\mathbf{m} = \bigg\{ \sum_{t=1}^n(\boldsymbol{\theta}_c^{(0)\mathsf{T}}\mathbf{c}_t)^2 \mathbf{B}_t\trans\mathbf{B}_t + \sigma^{2(0)} \lambda_{\mu_s}\bGamma\bigg\}^{-1} \sum_{t=1}^n(\boldsymbol{\theta}_c^{(0)\mathsf{T}}\mathbf{c}_t)\mathbf{B}_t\trans(\mathbf{z}_t - \mathbf{B}_t \mathbf{\Theta}^{(0)} \widehat{\balpha}_t)
\]
and 
\[
	\mathbf{A} = \sum_{t=1}^n(\boldsymbol{\theta}_c^{(0)\mathsf{T}}\mathbf{c}_t)^2 \mathbf{B}_t\trans\mathbf{B}_t + \sigma^{2(0)} \lambda_{\mu_s}\bGamma.
\]
We propose to use the gradient descent algorithm on sphere sub-manifold to obtain the minimizer of \eqref{sphereopti}. 
The gradient descent algorithm for a sub-manifold includes four iterative steps:
\begin{enumerate}
	\item[i] calculate the negative gradient of the objective function in the Euclidean space without any constraint;
	\item[ii] project the obtained negative gradient function onto the tangent space of manifold;
	\item[iii] evaluate the updating value along the direction of the projected negative gradient in step ii with a given step size;
	\item[iv] retract the calculated value in step iii back to the manifold structure.
\end{enumerate}
The step size in the above step iii can be determined using the Armijo backtracking method \citep[see, for example, Chapter 4.2 of][]{absil2009optimization}.

\begin{algorithm}[!h]
	\begin{algorithmic}
		\REQUIRE 
		Scalars $\beta, \gamma \in (0,1)$ and 
		initialization $\widehat{\btheta}_b^{(0)}$.
		\FOR {$k=1,2,\dots$}
		\STATE i) Compute $\boldsymbol{\eta}_k = -2\big\{\bA(\widehat{\btheta}_b^{(k-1)}  - \mathbf{m}) - \widehat{\btheta}_b^{(k-1)}  \big( \widehat{\btheta}_b^{(k-1)} \big)\trans\bA(\widehat{\btheta}_b^{(k-1)}  - \mathbf{m})\big\}$.
		\STATE ii) Find the smallest integer $n\ge 0$ such that
		$f\{R_{\widehat{\btheta}_b^{(k-1)} }(\beta^n\boldsymbol{\eta}_k)\} 
		\le f(\widehat{\btheta}_b^{(k-1)} ) - \gamma  \beta^n \boldsymbol{\eta}_k\trans \boldsymbol{\eta}_k $, where $R_{\widehat{\btheta}_b^{(k-1)} }(\beta^n\boldsymbol{\eta}_k) = (\widehat{\btheta}_b^{(k-1)} +  \beta^n \boldsymbol{\eta}_k) \big/ \Vert\widehat{\btheta}_b^{(k-1)}  + \beta^n \boldsymbol{\eta}_k\Vert$.
		\STATE iii) $\widehat{\btheta}_b^{(k)}  = R_{\widehat{\btheta}_b^{(k-1)} }(\beta^n\boldsymbol{\eta}_k)$.
		\STATE iv) Repeat until $\Vert\widehat{\btheta}_b^{(k)}  - \widehat{\btheta}_b^{(k-1)} \Vert$ is small enough.
		\ENDFOR 
		\RETURN $\widehat{\btheta}_b = \widehat{\btheta}_b^{(k)}$.
	\end{algorithmic}
	\caption{Gradient decent algorithm to update $\btheta_b$. \label{alg:armijo}}
\end{algorithm}

Beginning with a given initial value $\widehat{\btheta}_b^{(0)}$, we update $\widehat{\btheta}_b^{(k)}$ from $\widehat{\btheta}_b^{(k-1)}$ at the $k$-th iteration, $k=1,2,\dots$, until the convergence condition is satisfied. 
To be specific, at the $k$-th iteration, we first calculate the negative gradient of the objective function in \eqref{sphereopti}, which is explicit as
\[
	-\frac{\partial f(\btheta_b)}{\partial \btheta_b} = -2 \bA (\btheta_b - \mathbf{m}).
\]
Let $\boldsymbol{\eta}_k$ denote the projected negative gradient function on the tangent space of sphere. According to \cite{absil2009optimization}, it can be shown that
\[
	\boldsymbol{\eta}_k = \mathrm{Proj} \bigg\{- \frac{\partial f(\btheta_b)}{\partial \btheta_b}\bigg\} =  -2 \big\{\bA(\widehat{\btheta}_b^{(k-1)}  - \mathbf{m}) - \widehat{\btheta}_b^{(k-1)}  \big( \widehat{\btheta}_b^{(k-1)} \big) \trans \bA(\widehat{\btheta}_b^{(k-1)}  - \mathbf{m})\big\}.
\]
Afterwards we evaluate the updating value along the direction of the projected negative gradient $\boldsymbol{\eta}_k$ with the step size determined by the Armijo backtracking method \citep[see, for example, Chapter 4.2 of][]{absil2009optimization}.
In particular, we let $\beta, \gamma \in (0,1)$ be some pre-given parameters and Armijo backtracking method parametrizes the step size as $\beta^n$ with $n \geq 0$. 
The exponent $n \ge 0$ is the smallest integer  such that
\[
	f\{R_{\widehat{\btheta}_b^{(k-1)} }(\beta^n\boldsymbol{\eta}_k)\}
	\le f(\widehat{\btheta}_b^{(k-1)} ) - \gamma  \beta^n \|\boldsymbol{\eta}_k \|_2^2
\]
is satisfied, where $R_{\widehat{\btheta}_b^{(k-1)}}(\cdot)$ is the retraction function.
For sphere, the retraction function can be chosen as the normalization \citep{absil2009optimization}, i.e., 
\[
	R_{\widehat{\btheta}_b^{(k-1)}}(\beta^n \boldsymbol{\eta}_k) = \frac{\widehat{\btheta}_b^{(k-1)} + \beta^n \boldsymbol{\eta}_k}{\Vert \widehat{\btheta}_b^{(k-1)} + \beta^n \boldsymbol{\eta}_k\Vert}.
\]
Finally, after determining $n$, we retract $\widehat{\btheta}_b^{(k-1)} + \beta^n \boldsymbol{\eta}_k$ back to the sphere as the updated $k$-th iteration, i.e., 
\[
	\widehat{\btheta}_b^{(k)}  = R_{\widehat{\btheta}_b^{(k-1)} }(\beta^n\boldsymbol{\eta}_k).
\]

The above discussion leads to Algorithm \ref{alg:armijo} to summarize the gradient decent method on the sphere sub-manifold.

\subsection{Updating $\btheta_c$}
For $\btheta_c$, the corresponding objective function that we need to minimize is 
$$ \frac{1}{\sigma^{2(0)}}  \sum_{t=1}^n 
(\bz_t - \bB_t \widehat{\btheta}_b \btheta_c\trans \bc_t - \bB_t \bTheta^{(0)} \widehat{\balpha}_t)\trans 
(\bz_t - \bB_t \widehat{\btheta}_b \btheta_c\trans \bc_t - \bB_t \bTheta^{(0)} \widehat{\balpha}_t) + 
\lambda_{\mu_t} \btheta_c \trans \bP \btheta_c.$$
We take the first derivative of 
the above objective function with respect to $\btheta_c$ to obtain
\begin{equation*}
\begin{aligned} 
\frac{2}{\sigma^{2(0)}} \sum_{t=1}^n (\bB_t \widehat{\btheta}_b \bc_t\trans) \trans (\bz_t - \bB_t \widehat{\btheta}_b \btheta_c\trans \bc_t - \bB_t \bTheta^{(0)} \widehat{\balpha}_t) + 2\lambda{\mu_t} \bP \btheta_c. 
\end{aligned}
\end{equation*}
By setting the aforementioned display to zero, we thus can update $\btheta_c$ as 
\[	
	\widehat{\boldsymbol{\theta}}_c = \bigg\{\sum_{t=1}^n(\bB_t \widehat{\boldsymbol{\theta}}_b\bc_t\trans)\trans(\bB_t \widehat{\boldsymbol{\theta}}_b\bc_t\trans) + \sigma^{2(0)} \lambda_{\mu_t} \bP\bigg\}^{-1}\sum_{t=1}^n(\bB_t \widehat{\boldsymbol{\theta}}_b\bc_t\trans)\trans(\bz_t - \bB_t \mathbf{\Theta}^{(0)}\widehat{\balpha}_t).
\]

\subsection{Updating $\sigma^2$}
For $\sigma^2$, the corresponding objective function that we need to minimize is 
\begin{equation*}
\begin{aligned} 
    &\frac{1}{\sigma^2} \sum_{t=1}^n 
    \big\{ (\bz_t - \bB_t \widehat\btheta_b \widehat\btheta_c\trans \bc_t - \bB_t \bTheta^{(0)} \widehat\balpha_t)\trans 
    (\bz_t - \bB_t \widehat\btheta_b \widehat\btheta_c\trans \bc_t - \bB_t \bTheta^{(0)} \widehat\balpha_t)\\
    & \qquad \qquad \qquad+ \mathrm{tr}(\bB_t \bTheta^{(0)} \widehat{\bSigma}_t {\bTheta^{(0)}}\trans\mathbf{B}_t\trans)\big\} + \sum_{t=1}^n  n_t \log \sigma^2.
\end{aligned}
\end{equation*}
We take the first derivative of the above objective function with respect to $\sigma^2$ to obtain
\begin{equation*}
\begin{aligned} 
& -\frac{1}{(\sigma^2)^2} \Big\{ (\bz_t - \bB_t \widehat\btheta_b \widehat\btheta_c\trans \bc_t - \bB_t \bTheta^{(0)} \widehat\balpha_t)\trans 
(\bz_t - \bB_t \widehat\btheta_b \widehat\btheta_c\trans \bc_t - \bB_t \bTheta^{(0)} \widehat\balpha_t) \\
&\qquad \qquad\qquad + \mathrm{tr}(\bB_t \bTheta^{(0)} \widehat{\bSigma}_t {\bTheta^{(0)}}\trans\mathbf{B}_t\trans)\Big\} + 
\frac{1}{\sigma^2} \sum_{t=1}^n  n_t,   
\end{aligned}
\end{equation*}
By setting the aforementioned display to zero, we thus can update $\sigma^2$ as 
\begin{equation*}
\begin{aligned}
	\widehat{\sigma}^2 &= \frac{1}{\sum_{t=1}^n n_t} \sum_{t=1}^n \Big\{(\bz_t - \bB_t \widehat{\boldsymbol{\theta}}_b\widehat{\boldsymbol{\theta}}_c\trans \bc_t - \bB_t \mathbf{\Theta}^{(0)}\widehat{\balpha}_t)\trans(\bz_t - \bB_t\widehat{\boldsymbol{\theta}}_b\widehat{\boldsymbol{\theta}}_c\trans \bc_t - \bB_t \mathbf{\Theta}^{(0)}\widehat{\balpha}_t) \\
	& \qquad \qquad \qquad \qquad + \mathrm{tr}(\bB_t\mathbf{\Theta}^{(0)}\widehat{\bSigma}_t \mathbf{\Theta}^{(0)\mathsf{T}}\bB_t\trans)\Big\}.
\end{aligned} 
\end{equation*}

\subsection{Updating $\sigma_j^2$}
For $\sigma_j^2$, the corresponding objective function that we need to minimize is
$$n \log \sigma_{j}^2 + \frac{1}{\sigma_{j}^2} \widehat{S}_j(\mathbf{k}^{(0)}_j).$$
We take the first derivative of the above objective function with respect to $\sigma_j^2$ and set it to zero to 
update $\sigma_j^2$ as
$\widehat{\sigma}_{j}^2 = {\widehat{S}_j(\mathbf{k}^{(0)}_j)}/{n}$, $j =1,\dots, J$, where $\widehat{S}_j(\mathbf{k}^{(0)}_j)$ is as \eqref{eqn:defSj} of the main paper with $\mathbf{k}^{(0)}_j$ (the $j$-th column of $\mathbf{K}^{(0)}$) plugged in.

\subsection{Updating $\mathbf{\Theta}$}
For $\mathbf{\Theta}$, we first update the columns of $\mathbf{\Theta} = (\boldsymbol{\theta}_1,\dots,\boldsymbol{\theta}_J)$ sequentially. 
Updating $\btheta_j$ is equivalent to finding the minimizer of the following objective function:
$$\sum_{t=1}^n \Big\Vert \bz_t - \bB_t \widehat{\btheta}_b \widehat{\btheta}_c\trans \bc_t - \sum_{j' \neq j} \bB_t \btheta_{j'} \widehat{\alpha}_{j',t} - \bB_t \btheta_j \widehat{\alpha}_{j,t} \Big\Vert^2  + \sum_{t=1}^n \mathrm{tr}(\bB_t \bTheta \widehat{\bSigma}_t\bTheta\trans \bB_t\trans) + \widehat{\sigma}^2 \lambda_{pc} \btheta_j\trans \bGamma \btheta_j $$
We take the first derivative of the above objective function with respect to $\btheta_j$ to obtain
\begin{equation*}
\begin{aligned}
    &-2\sum_{t=1}^n \widehat{\alpha}_{j,t}\bB_t\trans \left(\bz_t - \bB_t \widehat{\btheta}_b \widehat{\btheta}_c\trans \bc_t - \sum_{j' \neq j} \bB_t \btheta_{j'} \widehat{\alpha}_{j',t} - \bB_t \btheta_j \widehat{\alpha}_{j,t}\right) \\
    &\quad\quad\quad\quad
    + 2\sum_{t=1}^n \big( \widehat{\bSigma}_{t,jj}\bB_t\trans \bB_t \btheta_j + \widehat{\bSigma}_{t,j'j} \bB_t\trans \bB_t \btheta_{j'} \big)+
    2\widehat{\sigma}^2 \lambda_{pc} \bGamma \btheta_j.
\end{aligned}
\end{equation*}
By setting the aforementioned display to zero, we thus can update $\btheta_j$ as 
\[
	\widehat{\boldsymbol{\theta}}_j = \bigg\{\sum_{t=1}^n (\widehat{\alpha}_{j,t}^2 + \widehat{\bSigma}_{t,jj})\bB_t\trans\bB_t + \widehat{\sigma}^2\lambda_{pc}\bGamma\bigg\}^{-1} 
	 \sum_{t=1}^n \bB_t\trans \Big\{ (\bz_t - \bB_t\widehat{\boldsymbol{\theta}}_b\widehat{\boldsymbol{\theta}}_c\trans \bc_t)\widehat{\alpha}_{j,t} - \sum_{j'\neq j}(\widehat{\alpha}_{j',t}\widehat{\alpha}_{j,t} + \widehat{\bSigma}_{t,j'j})\bB_t\widehat{\boldsymbol{\theta}}_{j'} \Big\}. 
\]

To guarantee the orthonormality of  $\widehat{\bTheta}$, we utilize the spectral decomposition of $\widehat{\bTheta}\widehat{\mathbf{H}}_J \widehat{\bTheta}\trans$, where $\widehat{\mathbf{H}}_J = \mathrm{diag}(\widehat{\sigma}_1^2, \dots, \widehat{\sigma}_J^2)$. 
In particular, let $\widehat{\bTheta}\widehat{\mathbf{H}}_J \widehat{\bTheta}\trans = \widetilde{\mathbf{Q}} \widetilde{\mathbf{D}} \widetilde{\mathbf{Q}}\trans$,
where $\widetilde{\mathbf{Q}}$ is orthonormal and $ \widetilde{\mathbf{D}}$ is a diagonal matrix with decreasing diagonal elements.
We then replace $\widehat{\bTheta}$ and $\widehat{\mathbf{H}}_J$ by $\widetilde{\mathbf{Q}}$ and $\widetilde{\mathbf{D}}$, respectively.
Furthermore, we replace $\widehat{\balpha}_t$ with $\widetilde{\mathbf{Q}}\trans \widehat{\mathbf{\Theta}} \widehat{\balpha}_t$, and such transformation preserves the variance of $\widehat{\mathbf{\Theta}} \widehat{\balpha}_t$.

\subsection{Updating $\mathbf{K}$}
For $\mathbf{K} = (\mathbf{k}_1,\dots,\mathbf{k}_J)$, the corresponding objective function that we need to minimize is
\begin{equation}\label{sup:ols}
	\sum_{j=1}^J\bigg\{- \log \left|\mathbf{M}_{j}\right| + \frac{1}{\widehat{\sigma}_j^2} \widehat{S}_j(\mathbf{k}_j)\bigg\},
\end{equation}
where the form of $\mathbf{M}_{j}$ is given in \eqref{eqn:precisionExpre} and $\widehat{S}_j(\mathbf{k}_j)$ is presented in \eqref{eqn:defSj} of the main paper.
Note that the value of $\log \left| \mathbf{M}_{j} \right|$ is invariant with the change of sample size $n$, while the second term in \eqref{sup:ols} is $n$-dependent. To simplify the computation, we use the second term $\sum_{j=1}^J\big\{ ({1}/{\widehat{\sigma}_j^2}) \widehat{S}_j(\mathbf{k}_j)\big\}$ to approximate \eqref{sup:ols}, which is separate for each column of $\mathbf{K}$. 
Since $\widehat{S}_j(\mathbf{k}_j) = (1,\mathbf{k}_j\trans) \widehat{\mathbf{D}}_j (1,\mathbf{k}_j\trans)\trans$ has a quadratic form with respect to $\mathbf{k}_j$, 
taking the first derivative of $\widehat{S}_j(\mathbf{k}_j)$ with respect to $\mathbf{k}_j$ and setting it to zero, we obtain $\widehat{\mathbf{k}}_j = (\widehat{\bD}_{pj})^{-1}\widehat{\mathbf{d}}_j$, $j=1,\dots,J$,
where $\widehat{\mathbf{d}}_j = (\widehat{D}_{1,2,j}, \dots, \widehat{D}_{1,p+1,j})\trans$ and $\widehat{\mathbf{D}}_{pj}$ is the bottom right $p\times p$ major submatrix of $\widehat{\bD}_j$.

\section{Uncertainty Quantification}\label{sec:uncertain}

We further conducted the semi-parametric bootstrap to obtain the bootstrap standard deviation (SD) surfaces for the principal component functions of the sFPC model. 
The procedure of semi-parametric bootstrap is summarized as follows. 
Let $\widehat{\Xi} =  \{\widehat{\boldsymbol{\theta}}_b, \widehat{\boldsymbol{\theta}}_c, \widehat{\mathbf{\Theta}}, 
\widehat{\mathbf{K}},
\widehat{\sigma}^2, \{\widehat{\sigma}_{j}^2\}_{j=1}^J \}$ be the estimated parameters of fitting the proposed sFPC model to the data set. 
Denote the fitting residuals of sFPC by $\{\widehat{\epsilon}_t(x_{t1},y_{t1}),\dots,\widehat{\epsilon}_t(x_{tn_t},y_{tn_t})\}$, $t=1,\dots,n$.
For the $b$-th replicate of bootstrap, we first generated the serial correlated FPC scores $\widehat{\alpha}_{j,t}^{(b)}$, $j=1,\dots, J$ and $t=1,\dots, n$, by the autoregressive model with the estimated coefficients $\widehat{k}_{i,j}$ and variance $\widehat{\sigma}_j^2$. 
We then generated the individual functions with the estimated coefficients of basis expansion as $\widehat{f}_t^{(b)} = \mathbf{b}_t(x,y)\trans\widehat{\btheta}_b \widehat{\btheta}_c\trans \bc_t + \sum_{j=1}^J \widehat{\alpha}_{j,t}^{(b)}\bb_j(x,y) \widehat{\btheta}_j$ for each time $t$. Afterwards, we generated the bootstrap data by adding the individual functions evaluated at the locations $\{(x_{t1},y_{t1}),\dots, (x_{tn_t},y_{tn_t})\}$ with measurement errors randomly sampled with replacement from the residuals $\{\widehat{\epsilon}_t(x_{t1},y_{t1}),\dots,\widehat{\epsilon}_t(x_{tn_t},y_{tn_t})\}$ for $t=1,\dots,n$. 
We fitted the proposed sFPC model to the $b$-th replicate of bootstrap data and obtained the estimated PC functions $\widehat{\phi}_j(x,y)^{(b)}$, for $j=1,\dots, J$. 
We repeated the above procedure $B$ times, i.e., with $b=1,\dots,B$, and gathered the estimated $j$-th PC functions $\widehat{\phi}_j(x,y)^{(b)}$, $b=1,\dots,B$, to obtain the bootstrap SD surface for each $j$.

\begin{figure}[ht]
    \centering
    \includegraphics[width=0.9\textwidth]{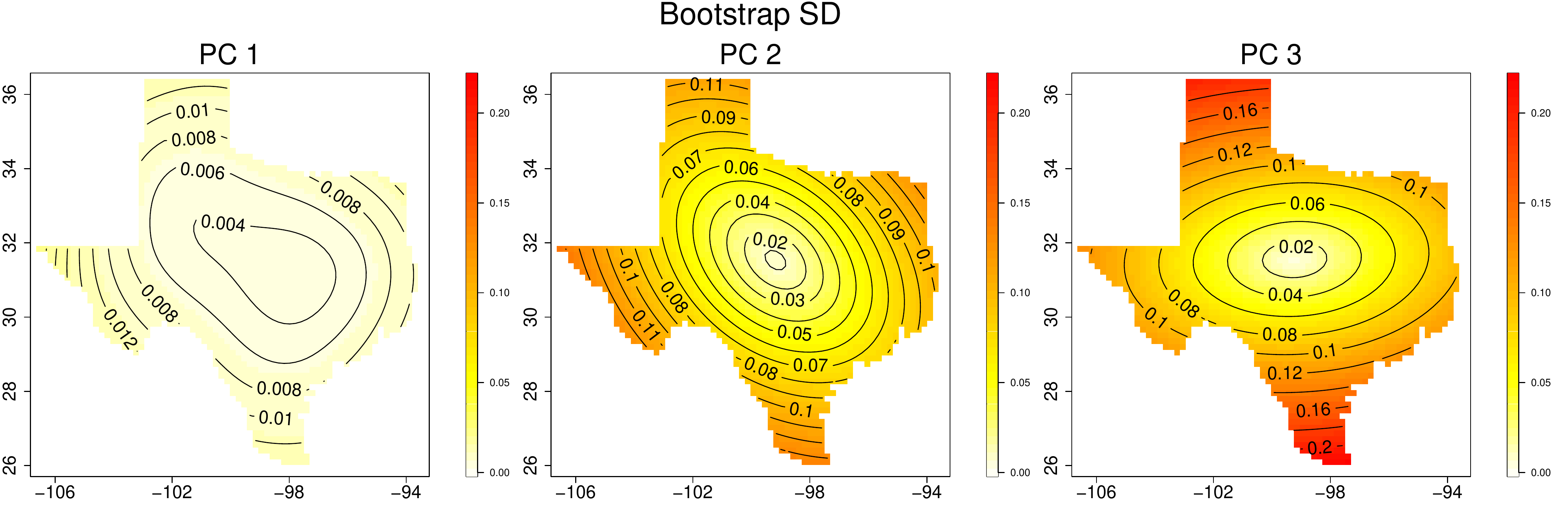}
    \caption{From left to right are the bootstrap SD surfaces of the first, second, and third  principal component functions, respectively, of the sFPC model for the real data analysis.}
    \label{fig:bootstrap}
\end{figure}

The corresponding results of the above procedure for our real data with $B = 100$ are depicted in Figure~\ref{fig:bootstrap}. 
We observe that all shapes of the bootstrap SD surfaces of PCs behave with lower values in the central areas and higher values in the boundary regions. 
This phenomenon may be explained by the fact that there are fewer stations in the boundary regions and more stations in the central areas.
Specific to each component, the values of bootstrap SD is up to $0.02$ for the first PC. 
Compared with the magnitude of the fitted first PC function, the relatively smaller SD values demonstrate the stable performance of sFPC on this leading component. 
On the other hand, the values of bootstrap SD are up to $0.14$ and $0.20$ for the second and third PCs, respectively.  
The larger SD values of these two components show that there are more uncertainty to estimate the second and third PC functions.

\spacingset{0.90}
\bibliographystyle{jasa}
\bibliography{reference}

\begin{thebibliography}{47}
\newcommand{\enquote}[1]{``#1''}
\expandafter\ifx\csname natexlab\endcsname\relax\def\natexlab#1{#1}\fi
\expandafter\ifx\csname url\endcsname\relax
  \def\url#1{{\tt #1}}\fi
\expandafter\ifx\csname urlprefix\endcsname\relax\def\urlprefix{URL }\fi

\bibitem[{Absil et~al.(2009)Absil, Mahony, and
  Sepulchre}]{absil2009optimization}
Absil, P.~A., Mahony, R., and Sepulchre, R.
\newblock {\em Optimization algorithms on matrix manifolds\/}.
\newblock Princeton, NJ: Princeton University Press (2009).

\bibitem[{Akaike(1974)}]{akaike1974new}
Akaike, H.
\newblock \enquote{A new look at the statistical model identification.}
\newblock {\em IEEE Transactions on Automatic Control\/}, 19(6):716--723
  (1974).

\bibitem[{Bosq(2000)}]{bosq2000linear}
Bosq, D.
\newblock {\em Linear processes in function spaces: Theory and applications\/}.
\newblock New York: Springer (2000).

\bibitem[{Box et~al.(2015)Box, Jenkins, Reinsel, and Ljung}]{box2015time}
Box, G.~E., Jenkins, G.~M., Reinsel, G.~C., and Ljung, G.~M.
\newblock {\em Time series analysis: Forecasting and control\/}.
\newblock Hoboken: John Wiley \& Sons (2015).

\bibitem[{Cabrera and Schulz(2017)}]{cabrera2017forecasting}
Cabrera, B.~L. and Schulz, F.
\newblock \enquote{Forecasting generalized quantiles of electricity demand: A
  functional data approach.}
\newblock {\em Journal of the American Statistical Association\/},
  112(517):127--136 (2017).

\bibitem[{Chen and Jiang(2017)}]{chen2017multi}
Chen, L.-H. and Jiang, C.-R.
\newblock \enquote{Multi-dimensional functional principal component analysis.}
\newblock {\em Statistics and Computing\/}, 27(5):1181--1192 (2017).

\bibitem[{Cipra and Romera(1997)}]{Cipra1997KalmanFW}
Cipra, T. and Romera, R.
\newblock \enquote{Kalman filter with outliers and missing observations.}
\newblock {\em Test\/}, 6:379--395 (1997).

\bibitem[{de~Boor(1978)}]{de1978practical}
de~Boor, C.
\newblock {\em A practical guide to splines\/}.
\newblock New York: Springer-Verlag (1978).

\bibitem[{Dempster et~al.(1977)Dempster, Laird, and
  Rubin}]{dempster1977maximum}
Dempster, A.~P., Laird, N.~M., and Rubin, D.~B.
\newblock \enquote{Maximum likelihood from incomplete data via the EM
  algorithm.}
\newblock {\em Journal of the Royal Statistical Society: Series B
  (Methodological)\/}, 39(1):1--22 (1977).

\bibitem[{Ding et~al.(2022)Ding, He, Jones, and Huang}]{ding2022functional}
Ding, F., He, S., Jones, D.~E., and Huang, J.~Z.
\newblock \enquote{Functional PCA with covariate-dependent mean and covariance
  structure.}
\newblock {\em Technometrics\/}, 64(3):335--345 (2022).

\bibitem[{Durbin and Koopman(2012)}]{durbin2012time}
Durbin, J. and Koopman, S.~J.
\newblock {\em Time series analysis by state space methods\/}.
\newblock Oxford: Oxford University Press, 2nd edition (2012).

\bibitem[{Golub and Van~Loan(2013)}]{golub2013matrix}
Golub, G.~H. and Van~Loan, C.~F.
\newblock {\em Matrix computations\/}.
\newblock Baltimore, MD: JHU press (2013).

\bibitem[{Hall et~al.(2006)Hall, M{\"u}ller, and Wang}]{hall2006properties}
Hall, P., M{\"u}ller, H.-G., and Wang, J.-L.
\newblock \enquote{Properties of principal component methods for functional and
  longitudinal data analysis.}
\newblock {\em The Annals of Statistics\/}, 34(3):1493--1517 (2006).

\bibitem[{Hansen et~al.(2006)Hansen, Sato, Ruedy, Lo, Lea, and
  Medina-Elizade}]{hansen2006global}
Hansen, J., Sato, M., Ruedy, R., Lo, K., Lea, D.~W., and Medina-Elizade, M.
\newblock \enquote{Global temperature change.}
\newblock {\em Proceedings of the National Academy of Sciences\/},
  103(39):14288--14293 (2006).

\bibitem[{Huang et~al.(2008)Huang, Shen, and Buja}]{huang2008functional}
Huang, J.~Z., Shen, H., and Buja, A.
\newblock \enquote{Functional principal components analysis via penalized rank
  one approximation.}
\newblock {\em Electronic Journal of Statistics\/}, 2:678--695 (2008).

\bibitem[{Hyndman and Shang(2009)}]{hyndman2009forecasting}
Hyndman, R.~J. and Shang, H.~L.
\newblock \enquote{Forecasting functional time series.}
\newblock {\em Journal of the Korean Statistical Society\/}, 38(3):199--211
  (2009).

\bibitem[{Hyndman and Ullah(2007)}]{hyndman2007robust}
Hyndman, R.~J. and Ullah, M.~S.
\newblock \enquote{Robust forecasting of mortality and fertility rates: A
  functional data approach.}
\newblock {\em Computational Statistics \& Data Analysis\/}, 51(10):4942--4956
  (2007).

\bibitem[{James et~al.(2000)James, Hastie, and Sugar}]{james2000principal}
James, G.~M., Hastie, T.~J., and Sugar, C.~A.
\newblock \enquote{Principal component models for sparse functional data.}
\newblock {\em Biometrika\/}, 87(3):587--602 (2000).

\bibitem[{Jones et~al.(1986)Jones, Wigley, and Wright}]{jones1986global}
Jones, P.~D., Wigley, T.~M., and Wright, P.~B.
\newblock \enquote{Global temperature variations between 1861 and 1984.}
\newblock {\em Nature\/}, 322(6078):430--434 (1986).

\bibitem[{Karhunen(1946)}]{karhunen1946spektraltheorie}
Karhunen, K.
\newblock \enquote{Zur spektraltheorie stochastischer prozesse.}
\newblock {\em Annales Academiae Scientiarum Fennicae. Series A. I,
  Mathematica\/}, 34:1--7 (1946).

\bibitem[{Kokoszka and Reimherr(2013)}]{kokoszka2013determining}
Kokoszka, P. and Reimherr, M.
\newblock \enquote{Determining the order of the functional autoregressive
  model.}
\newblock {\em Journal of Time Series Analysis\/}, 34(1):116--129 (2013).

\bibitem[{Lai and Schumaker(2007)}]{lai2007spline}
Lai, M.-J. and Schumaker, L.~L.
\newblock {\em Spline functions on triangulations\/}.
\newblock Cambridge: Cambridge University Press (2007).

\bibitem[{Li and Guan(2014)}]{li2014functional}
Li, Y. and Guan, Y.
\newblock \enquote{Functional principal component analysis of spatiotemporal
  point processes with applications in disease surveillance.}
\newblock {\em Journal of the American Statistical Association\/},
  109(507):1205--1215 (2014).

\bibitem[{Liu et~al.(2017)Liu, Ray, and Hooker}]{liu2017functional}
Liu, C., Ray, S., and Hooker, G.
\newblock \enquote{Functional principal component analysis of spatially
  correlated data.}
\newblock {\em Statistics and Computing\/}, 27(6):1639--1654 (2017).

\bibitem[{Liu et~al.(2016)Liu, Guillas, and Lai}]{liu2016efficient}
Liu, X., Guillas, S., and Lai, M.-J.
\newblock \enquote{Efficient spatial modeling using the SPDE approach with
  bivariate splines.}
\newblock {\em Journal of Computational and Graphical Statistics\/},
  25(4):1176--1194 (2016).

\bibitem[{Lo{\`e}ve(1946)}]{loeve1946fonctions}
Lo{\`e}ve, M.
\newblock \enquote{Fonctions al{\'e}atoires {\`a} d{\'e}composition orthogonale
  exponentielle.}
\newblock {\em La Revue Scientifique\/}, 84:159--162 (1946).

\bibitem[{Lorentz(1986)}]{lorentz2013bernstein}
Lorentz, G.~G.
\newblock {\em Bernstein polynomials\/}.
\newblock New York: Chelsea Publishing Co., 2nd edition (1986).

\bibitem[{Menne et~al.(2009)Menne, Williams~Jr, and Vose}]{USHCN2009us}
Menne, M.~J., Williams~Jr, C.~N., and Vose, R.~S.
\newblock \enquote{The US Historical Climatology Network monthly temperature
  data, version 2.}
\newblock {\em Bulletin of the American Meteorological Society\/},
  90(7):993--1008 (2009).

\bibitem[{Mercer(1909)}]{mercer1909}
Mercer, J.
\newblock \enquote{Functions of positive and negative type, and their
  connection the theory of integral equations.}
\newblock {\em Philosophical Transactions of the Royal Society of London.
  Series A, Containing Papers of a Mathematical or Physical Character\/},
  209:415--446 (1909).

\bibitem[{Nelder and Mead(1965)}]{Nelder1965simplex}
Nelder, J.~A. and Mead, R.
\newblock \enquote{A simplex method for function minimization.}
\newblock {\em The Computer Journal\/}, 7(4):308--313 (1965).

\bibitem[{Ramsay and Silverman(2005)}]{ramsay2006functional}
Ramsay, J.~O. and Silverman, B.~W.
\newblock {\em Functional data analysis\/}.
\newblock New York: Springer Science \& Business Media, 2nd edition (2005).

\bibitem[{Rice and Wu(2001)}]{rice2001nonparametric}
Rice, J.~A. and Wu, C.~O.
\newblock \enquote{Nonparametric mixed effects models for unequally sampled
  noisy curves.}
\newblock {\em Biometrics\/}, 57(1):253--259 (2001).

\bibitem[{Ruppert et~al.(2003)Ruppert, Wand, and
  Carroll}]{ruppert2003semiparametric}
Ruppert, D., Wand, M.~P., and Carroll, R.~J.
\newblock {\em Semiparametric regression\/}.
\newblock Cambridge: Cambridge University Press (2003).

\bibitem[{Schwarz(1978)}]{schwarz1978estimating}
Schwarz, G.
\newblock \enquote{Estimating the dimension of a model.}
\newblock {\em The Annals of Statistics\/}, 6(2):461--464 (1978).

\bibitem[{Shang and Hyndman(2017)}]{shang2017grouped}
Shang, H.~L. and Hyndman, R.~J.
\newblock \enquote{Grouped functional time series forecasting: An application
  to age-specific mortality rates.}
\newblock {\em Journal of Computational and Graphical Statistics\/},
  26(2):330--343 (2017).

\bibitem[{Shen(2009)}]{shen2009modeling}
Shen, H.
\newblock \enquote{On modeling and forecasting time series of smooth curves.}
\newblock {\em Technometrics\/}, 51(3):227--238 (2009).

\bibitem[{Shen and Huang(2008)}]{shen2008interday}
Shen, H. and Huang, J.~Z.
\newblock \enquote{Interday forecasting and intraday updating of call center
  arrivals.}
\newblock {\em Manufacturing \& Service Operations Management\/},
  10(3):391--410 (2008).

\bibitem[{Shi et~al.(2022)Shi, Yang, Wang, Ma, Beg, Pei, and Cao}]{shi2022two}
Shi, H., Yang, Y., Wang, L., Ma, D., Beg, M.~F., Pei, J., and Cao, J.
\newblock \enquote{Two-dimensional functional principal component analysis for
  image feature extraction.}
\newblock {\em Journal of Computational and Graphical Statistics\/},
  31(4):1127--1140 (2022).

\bibitem[{Staniswalis and Lee(1998)}]{staniswalis1998nonparametric}
Staniswalis, J.~G. and Lee, J.~J.
\newblock \enquote{Nonparametric regression analysis of longitudinal data.}
\newblock {\em Journal of the American Statistical Association\/},
  93(444):1403--1418 (1998).

\bibitem[{Wang et~al.(2020{\natexlab{a}})Wang, Wang, Lai, and
  Gao}]{Wang2019Triangulation}
Wang, L., Wang, G., Lai, M.-J., and Gao, L.
\newblock \enquote{Efficient estimation of partially linear models for data on
  complicated domains by bivariate penalized splines over triangulations.}
\newblock {\em Statistica Sinica\/}, 30:347--369 (2020{\natexlab{a}}).

\bibitem[{Wang et~al.(2020{\natexlab{b}})Wang, Wang, Wang, and
  Ogden}]{wang2020simultaneous}
Wang, Y., Wang, G., Wang, L., and Ogden, R.~T.
\newblock \enquote{Simultaneous confidence corridors for mean functions in
  functional data analysis of imaging data.}
\newblock {\em Biometrics\/}, 76(2):427--437 (2020{\natexlab{b}}).

\bibitem[{Yao et~al.(2005{\natexlab{a}})Yao, M{\"u}ller, and
  Wang}]{yao:muller:wang:05}
Yao, F., M{\"u}ller, H.-G., and Wang, J.-L.
\newblock \enquote{Functional data analysis for sparse longitudinal data.}
\newblock {\em Journal of the American Statistical Association\/}, 100:577--590
  (2005{\natexlab{a}}).

\bibitem[{Yao et~al.(2005{\natexlab{b}})Yao, M{\"u}ller, and
  Wang}]{yao2005functional}
---.
\newblock \enquote{Functional linear regression analysis for longitudinal
  data.}
\newblock {\em The Annals of Statistics\/}, 33(6):2873--2903
  (2005{\natexlab{b}}).

\bibitem[{Yu et~al.(2020)Yu, Wang, Wang, Liu, and Yang}]{Yu2019Triangulation}
Yu, S., Wang, G., Wang, L., Liu, C., and Yang, L.
\newblock \enquote{Estimation and inference for generalized geoadditive
  models.}
\newblock {\em Journal of the American Statistical Association\/},
  115(530):761--774 (2020).

\bibitem[{Zhou et~al.(2008)Zhou, Huang, and Carroll}]{zhou2008joint}
Zhou, L., Huang, J.~Z., and Carroll, R.
\newblock \enquote{Joint modelling of paired sparse functional data using
  principal components.}
\newblock {\em Biometrika\/}, 95(3):601--619 (2008).

\bibitem[{Zhou and Pan(2014{\natexlab{a}})}]{Zhou2014Principal}
Zhou, L. and Pan, H.
\newblock \enquote{Principal component analysis of two-dimensional functional
  data.}
\newblock {\em Journal of Computational and Graphical Statistics\/},
  23(3):779--801 (2014{\natexlab{a}}).

\bibitem[{Zhou and Pan(2014{\natexlab{b}})}]{zhou2014smoothing}
---.
\newblock \enquote{Smoothing noisy data for irregular regions using penalized
  bivariate splines on triangulations.}
\newblock {\em Computational Statistics\/}, 29(1):263--281
  (2014{\natexlab{b}}).

\end{thebibliography}

\end{document}